\documentclass[12pt]{article}
\usepackage{color,graphicx}

\begin{document}

\title{Recalculation of Pion Compton Scattering in
Perturbative QCD}
\author{Ding-fang Zeng\\
\small Department of Physics, Peking University,
Beijing 100871, China\\
Bo-Qiang Ma\\
\small CCAST (World Laboratory), P.O. Box 8730, Beijing 100080,
China,\\
\small and Department of Physics, Peking University, Beijing
100871, China\footnote{Mailing address.}}


\date{\today }
\maketitle

\begin{abstract}
\begin{quote}
We recalculated pion virtual Compton scattering in perturbative
QCD in this paper. Our calculation avoids some deficiencies in
existing literatures, and treats real Compton scattering as a
limit case in which the mass of the virtual photon equals to zero.
Expressions of the hard scattering amplitudes from 10 independent
diagrams are given explicitly in the text. By comparing the
effects of different distribution amplitudes on the physical
observables, we studied the self-consistence of pQCD calculation
of this problem.
\end{quote}
\end{abstract}

\section{Introduction}

Pion virtual Compton scattering (VCS)
$\gamma^*\pi^-\rightarrow\gamma\pi^-$ via the reaction
$e\pi^-\rightarrow e\gamma\pi^-$ is observed by SELEX
Collaboration at Fermi Lab \cite{selex} for the first time.
Although in the current available kinematical region, the process
can not be predicted precisely in perturbative QCD (pQCD), it is
possible to observe such processes in the pQCD-applicable region
with the quick development of experimental techniques. Therefore
it is meaningful to check the pQCD prediction of this process.

For this problem, Tamazouzt \cite{Tama} calculated a very similar
process, $\gamma\pi^\pm\rightarrow\gamma^*\pi^\pm$; Maina and
Torrasso \cite{Maina} calculated it directly and treated the
singular integration appearing in it carefully; Li and Coriano
\cite{Hnli} calculated it by a rather different way. One common
problem existing in \cite{Tama} and \cite{Hnli} is: the authors
only directly calculated 5 diagrams contributing to the
unintegrated amplitude but gave no prescriptions about how to get
the other 15 ones, please see fig.\ref{Feynman-diagram} and
captions there. Ref.\cite{Maina} did not give its expressions for
the unintegrated amplitude but claimed consistence with
\cite{Tama}. As was pointed out by the author of \cite{Maina}, the
numerical treatments of \cite{Tama} have some defaults. However,
in literature \cite{Maina}, features that deserve further
investigations still exist after the revision. For example, a
nearly jumping change happened to the phase of $M_{LR}$, please
see figure 5 of \cite{Maina}.

\begin{figure}
\begin{center}
\scalebox{1}{\includegraphics*[113pt,578pt][451pt,767pt]{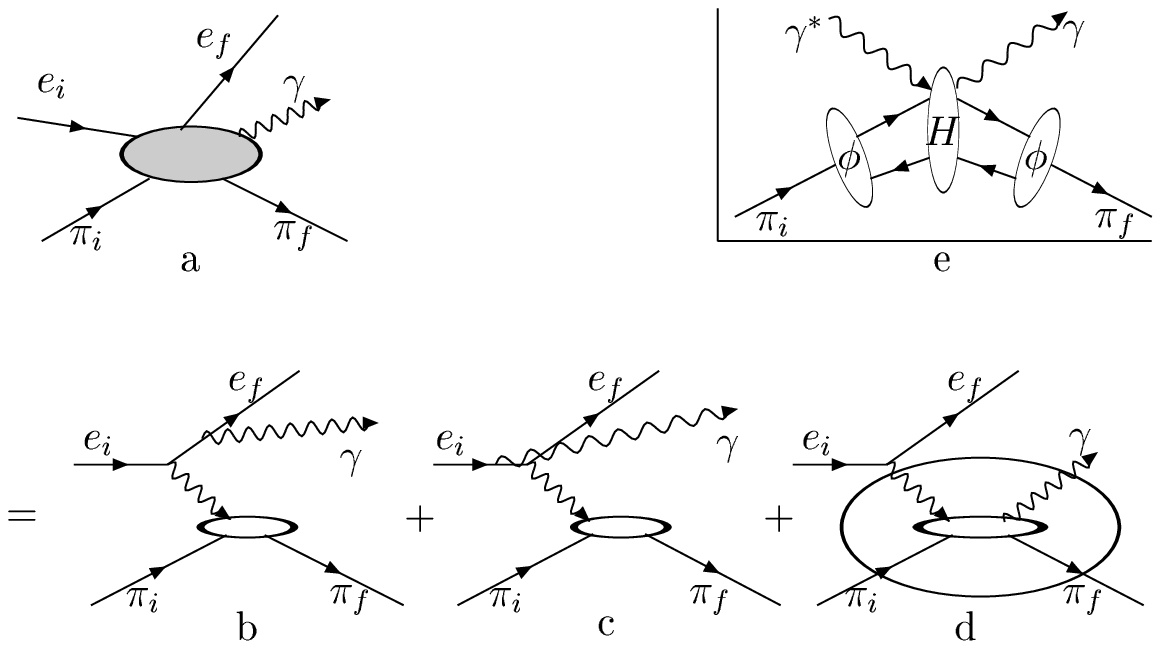}}
\begin{minipage}[b]{0.85\textwidth} \caption{\small
Reaction (a) $e\pi^\pm\rightarrow e\gamma\pi^\pm$ can proceed
both through (b)\&(c) Bethe-Heitler process and through (d)
virtual Compton scattering. In experiments, the two kinds of
process can not be separately detected. When the pion momentum
change is large enough in the process, the amplitude $M$ of the
process (d) $\gamma^*\pi^\pm\rightarrow\gamma\pi^\pm$ can be
factorized as (e) $M=\int dxdy\phi(x)H(x,y)\phi(y)$, where $H$
can be computed perturbatively, please see
fig.\ref{Feynman-diagram}. } \label{fig1}
\end{minipage}
\end{center}
\end{figure}

Because of these questions, we decide to recalculate this problem
in this paper. It will be shown that, all the deficiencies in the
literatures will not exist in our recalculated results.

\section{Factorization Theorem of pQCD for Pion VCS }\label{factor-sec}

As indicated in the caption of fig.1, a complete physical process
$e\pi^-\rightarrow e\gamma\pi^-$ can take place through two ways,
Bethe-Heitler process and virtual Compton scattering. But we will
not calculate such a complete process in this paper, please refer
to \cite{Maina} and \cite{GZh}. We will concentrate on the
sub-process $\gamma^*\pi^\pm\rightarrow\gamma\pi^\pm$ and use $M$
denoting the amplitude of it. Besides its inclusion of real
Compton scattering as a limit case, it is also meaningful in the
preparation for the calculation of the complete process
$e\pi^-\rightarrow e\gamma\pi^-$.

Factorization theorem states that for an exclusive process
\cite{Lepage} such as pion VCS
$\gamma^*\pi^\pm\rightarrow\gamma\pi^\pm$, the amplitude of it
can be written as the following convolution formula,
\begin{equation}
M(p;\epsilon q\rightarrow p^\prime;\epsilon^\prime q^\prime)=\int
dxdy\phi(x,Q)H(xp;\epsilon,q\rightarrow yp^\prime;\epsilon^\prime,
q^\prime)\phi(y,Q)\; , \label{factor}
\end{equation}
where $p$ denotes the momentum of the incoming pion, $\epsilon$
and $q$ are the polarization vector and momentum  of the photon
respectively. $x$ denotes one of the momentum fraction of the
valence quarks in the incoming pion, and that of the other one
will be denoted by $\bar{x}=1-x$. The primed variables or $y$ are
associated with outgoing particles.

The pion distribution amplitude $\phi(x,Q)$ in eq.(\ref{factor})
absorbs the long-distance dynamics of $M$ and can be derived by
non-perturbative methods \cite{sr-dis1}. Appearing of $Q$ in it
indicates its evolution with the energy scale. In this paper,
instead of consideration of such evolution \cite{Hnli,Botts}, we
will study the following five phenomenological models and their
effects on the physics predictions \cite{CZ,MR,PP,HMS,HS},
\begin{eqnarray}
&&\phi_{as}=\sqrt{3}f_\pi x(1-x),\nonumber\\
&&\phi_{bhl}=1.4706\sqrt{3}f_\pi
x(1-x)\exp[-\frac{0.07043}{x(1-x)}],\nonumber\\
&&\phi_{cz}=5\sqrt{3}f_\pi x(1-x)(2x-1)^2,\nonumber\\
&&\phi_{hs}=8.8763\sqrt{3}f_{\pi}x(1-x)(2x-1)^2\exp[-\frac{0.07062}{x(1-x)}],
\nonumber\\
&&\phi_{p3}=\sqrt{3}f_\pi
x(1-x)[0.6016-4.659(2x-1)^2+15.52(2x-1)^4],\nonumber\\
\label{da}
\end{eqnarray}
with the pion decay constant $f_\pi=93$~MeV and the distribution
amplitudes normalized by $\int d x \phi(x)=\sqrt{3}f_{\pi}/6$,
please see fig.4 for their explicit shapes. From the figure, we
can see that, relative to $\phi_{as}$ and $\phi_{cz}$, the
distribution amplitudes $\phi_{bhl}$ and $\phi_{hs}$ suppress the
end point region deeply, while function $\phi_{p3}$ intensifies
both the near-end-point region and the center region.

\begin{center}
\begin{figure}
\hspace{2cm}\begin{minipage}[b]{0.8\textwidth}
\scalebox{0.9}{\includegraphics*[77pt,539pt][499pt,766pt]{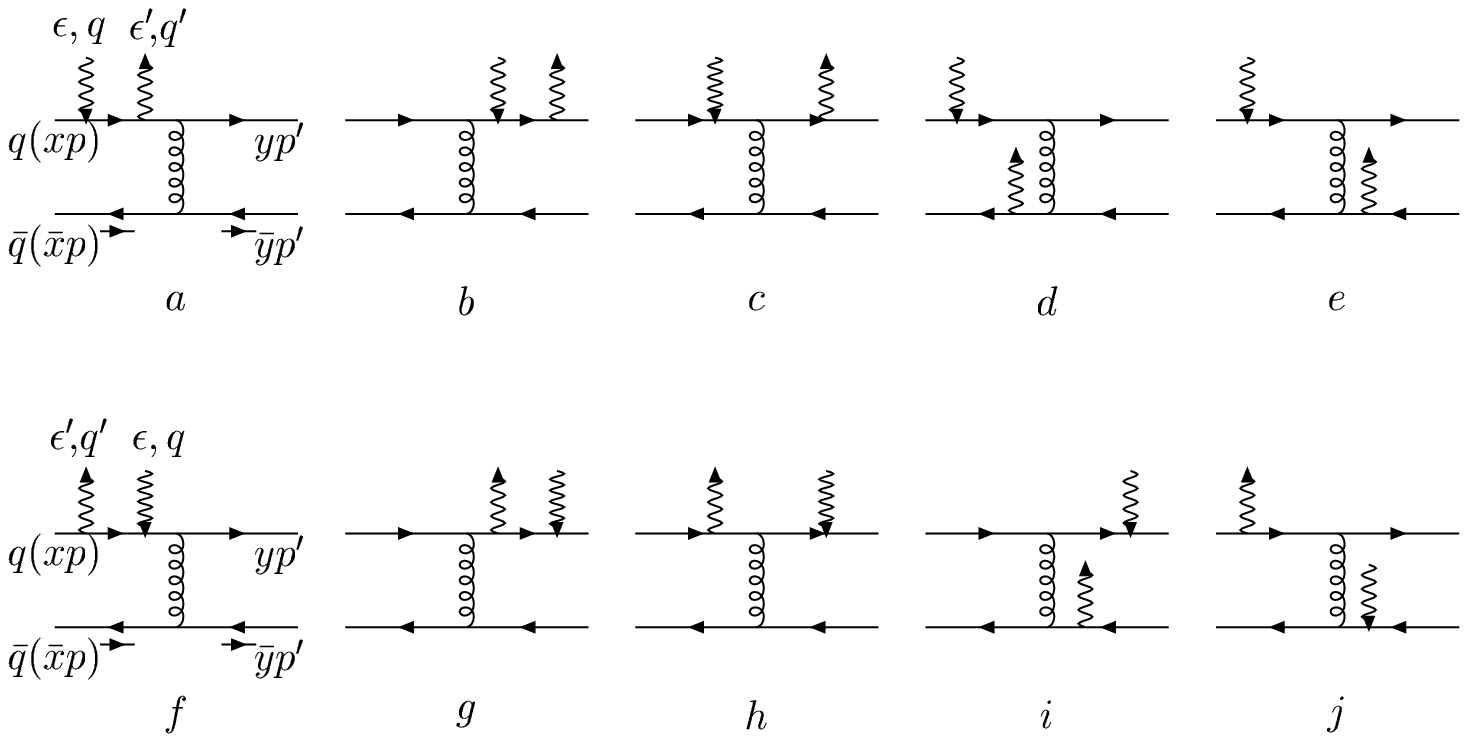}}
\caption{\small Unintegrated (hard) amplitude $H$ can be
calculated on the basis of these diagrams. Complete $H$ includes
other ten diagrams with the photons attaching to different quark
lines, and contributions of those diagrams to the full amplitude
$M$ are equal to the above ones except some charge factors.}
\label{Feynman-diagram}
\end{minipage}
\end{figure}
\end{center}

Contrary to $\phi(x)$, the hard amplitude $H(xp;\epsilon
q\rightarrow yp^\prime;\epsilon^\prime q^\prime)$ in
eq.(\ref{factor}) absorbs short-distance dynamics of the
amplitude, and can be calculated perturbatively on the basis of
the diagrams in fig.\ref{Feynman-diagram}. With leading Fock state
of $\pi^+$ ($\pi^-$ case is similar),
\begin{equation}
|\pi^+(p)\rangle
=\frac{1}{\sqrt{2}}\frac{1}{\sqrt{3}}\sum_{i}^{1,2,3}
\left[|u^i_\uparrow(xp)\rangle
|\bar{d}^{\bar{i}}_\downarrow(\bar{x}p)\rangle-
|u^i_\downarrow(xp)\rangle
|\bar{d}^{\bar{i}}_\uparrow(\bar{x}p)\rangle \right] ,
\end{equation}
where $i$ and $\bar{i}$ denote the color indices, $H$ can be
written as
\begin{eqnarray}
&&H(xp,\epsilon q\rightarrow yp^\prime,\epsilon^\prime
q^\prime)\sim\nonumber\\
&&\hspace{0.8cm}\sum_{diag.}\sum_{color}
\frac{1}{2}\frac{1}{3}\left[\frac{\bar{u}^{j}_\uparrow(yp^\prime)
(...\gamma^\mu t^a_{ji}...)u^i_\uparrow(xp) g_{\mu\nu}\delta^{ab}
\bar{v}^{\bar{i}}_\downarrow(\bar{x}p)(...\gamma^\nu
t^b_{\bar{i}\bar{j}}...)
v^{\bar{j}}_\downarrow(\bar{y}p^\prime)}{p_1^2p_2^2p_3^2}\right.
\nonumber\\
&&\hspace{2.5cm}\left.-\frac{\bar{u}^{j}_\downarrow(yp^\prime)
(...\gamma^\mu
t^a_{ji}...)u^{i}_\downarrow(xp)g_{\mu\nu}\delta^{ab}
\bar{v}^{\bar{i}}_\uparrow(\bar{x}p)(...\gamma^\nu
t^b_{\bar{i}\bar{j}} ...)
v^{\bar{j}}_\uparrow(\bar{y}p^\prime)}{p_1^2p_2^2p_3^2}\right],
\nonumber\\
\label{hdpt}
\end{eqnarray}
with the coupling constants $\alpha_e$, $\alpha_s$ and the charge
factor $e_u^2$ or $e_ue_{\bar{d}}$ suppressed for the moment. In
eq.(\ref{hdpt}), $p_1$, $p_2$ and $p_3$ denote the momentum
transferred through the two fermion and one gluon propagators;
explicit expressions of $(...\gamma^\mu t^a...)$s depend on the
details of diagrams. Using identity of $SU(3)$ color group,
\begin{equation}
t^a_{ij}t^a_{kl}=\frac{1}{2}(\delta_{il}\delta_{jk}-\frac{1}{3}\delta_{ij}\delta_{kl}),
\end{equation}
and some trace making trick, please see \cite{Field},
eq.(\ref{hdpt}) can be transformed into the following form
\begin{eqnarray}
&&H(x,\epsilon q\rightarrow y,\epsilon^\prime
q^\prime)=\sum_{diag.}H^{(diag.)}_{\epsilon\epsilon^\prime}\nonumber\\
&&\hspace{3.2cm}=\frac{2}{3}\sqrt{x\bar{x}y\bar{y}}\sum_{diag.}
Tr[(\gamma^\mu...)_{diag.}\gamma^5p\!\!\!/(\gamma_\mu...)_{diag.}
\gamma^5p^\prime\!\!\!\!/]. \label{sumdiagram}
\end{eqnarray}
By the usual convention, distribution amplitude $\phi(x)\phi(y)$
will absorb the $\sqrt{x\bar{x}y\bar{y}}$ factor, so it will not
be included in our later expressions for
$H_{\epsilon\epsilon^\prime}$.

\section{Unintegrated Amplitude of
$\gamma^*\pi^\pm\rightarrow\gamma\pi^\pm$}\label{hdpt-sec}

In the center-of-momentum frame of outgoing particles $\gamma$
and $\pi^\pm$, please refer to \cite{GZh}, we write all the
relevant kinematical variables as follows,
\begin{eqnarray}
&&p^{\prime\mu}=\frac{\omega+p}{2}(1,-\sin\theta,0,-\cos\theta),\nonumber\\
&&q^{\prime\mu}=\frac{\omega+p}{2}(1,\sin\theta,0,\cos\theta),\nonumber\\
&&p^\mu=(p,0,0,-p),\ \ q^\mu=(\omega,0,0,p),\label{kinematics}\\
&&\epsilon_R^\mu=\frac{1}{\sqrt{2}}(0,-1,-i,0), \ \
\epsilon_R^{\prime\mu}=\frac{1}{\sqrt{2}}(0,-\cos\theta,-i,\sin\theta),\nonumber\\
&&\epsilon_{L}^\mu=\frac{1}{\sqrt{2}}(0,1,-i,0),\nonumber\\
&&\epsilon_+^\mu=\frac{1}{\sqrt{2}}(1,0,0,1) . \label{photonpola}
\end{eqnarray}

Obviously, $\theta$ denotes the scattering angle of the process.
According to parity invariance and gauge invariance, we only need
to calculate three helicity amplitudes for the purpose of
computing the amplitude of the complete physical process
$e\pi^\pm\rightarrow e\gamma\pi^\pm$, please see \cite{Maina} and
\cite{GZh}. In order to compare with \cite{Tama}, we choose to
calculate $H_{RR}$, $H_{LR}$ and $H_{+R}$, while obtain the other
five by the following relations,
\begin{eqnarray}
&&H_{LL}=H_{RR},\ \ H_{RL}=H_{LR},\ \ H_{+L}=H_{+R};\nonumber\\
&&H_{-R}=-v^{-1}H_{+R},\ \ H_{-L}=-v^{-1}H_{+L}.\label{parity}
\end{eqnarray}

After introducing the abbreviation $c=\cos\frac{\theta}{2}$,
$s=\sin\frac{\theta}{2}$, $S=(\omega+p)^2$, $v=q^\mu q_\mu/S$ and
$\bar{v}=1-v$, we can write all the diagrams contributing to $H$
in more economical forms. The results are shown in table 1-3,
where we have transformed the diagrams with two propagators
potentially on shell into some equivalent forms in which only one
of them can go on shell by the following relation,
\begin{equation}
\frac{f(x,y)}{(x-a(y)+i\epsilon)(x-b(y)+i\epsilon)}
=\frac{f(x,y)}{a(y)-b(y)}
\left(\frac{1}{x-a(y)+i\epsilon}-\frac{1}{x-b(y)+i\epsilon}\right).\label{singular}
\end{equation}
This is very important in the numerical integration. Considering
the fact that each diagram of fig.\ref{Feynman-diagram} has a
companion with its photons attached to different quark lines, we
include a charge factor for each of the diagrams in the table,
where $\frac{5}{9}=e_u^2+e_{\bar{d}}^2,
\frac{4}{9}=2e_ue_{\bar{d}}$.
\begin{center}{Table 1 $H^{(diag.)}_{RR}(\frac{1}{S}\alpha_e\alpha_s$)}\\%
\[\begin{array}{cccc}\hline\hline%
{a:}&{\frac{5}{9}\frac{4c^2s^{-2}\bar{v}^{-1}}{\bar{x}\bar{y}(x-a-i\epsilon)}}
&{f:}&{\frac{5}{9}\frac{4s^{-2}\bar{v}^{-1}}{x\bar{x}\bar{y}}}\\%
{b:}&{\frac{5}{9}\frac{4c^2s^{-2}\bar{v}^{-1}}{\bar{x}\bar{y}y}}
&{g:}&{\frac{5}{9}\frac{4c^2s^{-2}\bar{v}^{-1}}{\bar{x}\bar{y}[(1-\bar{v}s^2)y-v]}}\\%
{c:}&{\frac{5}{9}\frac{-4c^2s^{-2}\bar{v}^{-1}}{\bar{x}y\bar{y}(x-a+i\epsilon)}}
&{h:}&{\frac{5}{9}\frac{-4s^{-2}\bar{v}^{-1}[1-(\bar{v}x+y)s^2+2\bar{v}xys^4]}
{[y(1-\bar{v}s^2)-v]x\bar{x}\bar{y}}}\\%
{d:}&{\frac{4}{9}\frac{4\bar{v}^{-1}[c^2+s^2(\bar{v}x+v)]}{c^2\bar{x}y}
(\frac{1}{x-a+i\epsilon}-\frac{1}{x-b+i\epsilon})}
&{i:}&{\frac{4}{9}\frac{-4c^2}{\bar{y}[y(1-\bar{v}s^2)-v](x-b+i\epsilon)}}\\%
{e:}&{0}
&{j:}&{\frac{4}{9}\frac{-4s^2\bar{v}^{-1}[y-v+\bar{v}\bar{x}(1-2ys^2)]}
{\bar{x}(1-ys^2)[y(1-\bar{v}s^2)-v](x-b+i\epsilon)}}\\%
\hline\hline{}&{
a=-\frac{v}{\bar{v}},b=\frac{y-v-y\bar{v}s^2}{\bar{v}(1-ys^2)}}&{}&{}
\end{array}
\]
\end{center}

\begin{center}{Table 2 $H^{(diag.)}_{LR}(\frac{1}{S}\alpha_e\alpha_s)$}\\%
\[\begin{array}{cccc}\hline\hline%
{a:}&{\frac{5}{9}\frac{4\bar{v}^{-1}}{\bar{x}\bar{y}(x-a+i\epsilon)}}
&{f:}&{0}\\%
{b:}&{\frac{5}{9}\frac{4\bar{v}^{-1}}{\bar{x}y\bar{y}}}
&{g:}&{\frac{5}{9}\frac{4v\bar{v}^{-1}}{\bar{x}\bar{y}[(1-\bar{v}s^2)y-v]}}\\%
{c:}&{0}
&{h:}&{\frac{5}{9}\frac{-4\bar{v}^{-1}[y+\bar{v}x(1-2ys^2)]}{x\bar{x}\bar{y}[(1-\bar{v}s^2)y-v]}}\\%
{d:}&{\frac{4}{9}\frac{4c^{-2}s^2}{y}(\frac{1}{x-a+i\epsilon}-\frac{1}{x-b+i\epsilon})}
&{i:}&{\frac{4}{9}\frac{-4c^2s^2y}{\bar{y}(1-ys^2)[(1-\bar{v}s^2)y-v](x-b+i\epsilon)}}\\%
{e:}&{\frac{4}{9}\frac{-4c^{-2}s^2}{y\bar{y}}(\frac{1}{x-a+i\epsilon}-\frac{1}{x-b+i\epsilon})}
&{j:}&{\frac{4}{9}\frac{-4s^2\bar{v}^{-1}[\bar{y}-\bar{v}\bar{x}(1-2ys^2)]}
{\bar{x}(1-ys^2)[(1-\bar{v}s^2)y-v](x-b+i\epsilon)}}
{}\\%
\hline\hline\end{array}
\]
\end{center}

\begin{center}{Table 3 $H^{(diag.)}_{+R}(\frac{1}{S}\alpha_e\alpha_s)$}\\%
\[\begin{array}{cccccccc}\hline\hline%
{a:}&{\frac{5}{9}\frac{8cs^{-1}}{\bar{x}\bar{y}}}
&{f:}&{\frac{5}{9}\frac{4c^{-1}s^{-1}\bar{v}^{-1}(1-2\bar{v}x)}{x\bar{x}\bar{y}}}\\%
{b:}&{\frac{5}{9}\frac{4cs^{-1}\bar{v}^{-1}}{\bar{x}y\bar{y}}}
&{g:}&{\frac{5}{9}\frac{4cs^{-1}v\bar{v}^{-1}}{\bar{x}\bar{y}[(1-\bar{v}s^2)y-v]}}\\%
{c:}&{\frac{5}{9}\frac{-4cs^{-1}}{\bar{x}y\bar{y}}}
&{h:}&{\frac{5}{9}\frac{-4cs^{-1}\bar{v}^{-1}y(1-2\bar{v}s^2x)}{x\bar{x}\bar{y}[y(1-\bar{v}s^2)-v]}}\\%
{d:}&{\frac{4}{9}\frac{-4c^{-1}s\bar{v}^{-1}[1-2\bar{v}s^2\bar{x}]}{\bar{x}(1-ys^2)(x-b+i\epsilon)}}
&{i:}&{\frac{4}{9}\frac{-4c^3sy}{\bar{y}(1-ys^2)[y(1-\bar{v}s^2)-v](x-b+i\epsilon)}}\\%
{e:}&{\frac{4}{9}\frac{4cs}{\bar{y}(1-ys^2)(x-b+i\epsilon)}}
&{j:}&{\frac{4}{9}\frac{-4cs\bar{v}^{-1}[v-y+2\bar{v}\bar{x}ys^2]}
{\bar{x}(1-ys^2)[y(1-\bar{v}s^2)-v](x-b+i\epsilon)}}\\%
\hline\hline\end{array}
\]
\end{center}

For $H^{(diag.)}_{RR}$ and $H^{(diag.)}_{LR}$, we compared our
expressions with those of \cite{Tama} in the $v\rightarrow0$ limit
(in \cite{Tama}, it is $R_b\rightarrow0$). Except our
consideration of 5 additional independent diagrams labelled $f \to
j$, all the other terms, labelled $a \to e$, coincide with
\cite{Tama}. For $H^{(diag.)}_{+R}$, in \cite{Tama}, it is
$H_{0R}$, because we employ the convention of \cite{Maina} for the
virtual photon polarization vector, which is different from
\cite{Tama}, our expression of it does not coincide with that of
\cite{Tama}. Our convention is very convenient for future
calculation of the complete process $e\pi^{\pm} \to e\gamma
\pi^{\pm}$.

In the case of $v=0$, by adding all the diagrams in table 2
together, we can get a rather simple expression for $H_{LR}$,
\begin{equation}
H_{LR}=-(\frac{2}{3}+\frac{1}{3})^2\frac{8c^{-2}s^2}{x(1-y)},
\end{equation}
So, in the $v\to 0$ limit, the amplitude $H_{LR}$ is a real
number, it has no imaginary part. About this point, literature
\cite{singular-of-hlr} give a general discussion. It should be
notified that \cite{Maina} has an error or misprint in giving its
expression for $H_{LR}|_{v\to0}$ as
$H_{LR}=C_0(e_1-e_2)^2(x-y)c^{-2}s^2$. Obviously, such an
unintegrated amplitude will give zero amplitude $M_{LR}$ in the
integration after multiplied by a symmetric factor
$\phi(x)\phi(y)$, please see eq.(\ref{factor}).

From table 1-3, we can see that, one by one, diagrams in the
second row of fig.{\ref{Feynman-diagram} do not equal to those of
the first row. From the numerical results of later sections, we
will be able to see that, as a total, the second row diagrams also
do not equal to those of the first row. So in this problem, the
number of independent diagrams is 10 instead of 5. Of course, the
total number is 20 as we indicate in the caption of
fig.\ref{Feynman-diagram}.

\section{Analytical Results of Electron VCS and Qualitative
Properties of Pion VCS}\label{sec-evcs}

From the aspect of experiencing VCS, unpolarized electrons and
pions are similar to each other,  so we can hope cross sections
of VCS on the unpolarized electrons and on pions have similar $v$
and $\theta$ dependence. Because electron has no internal
structure, its VCS cross sections can be get analytically. By the
same kinematical variables as those of pion VCS, neglecting the
mass of the electron, we can get the following expressions for
electron VCS
$\gamma^*_{\epsilon}e\rightarrow\gamma_{\epsilon^\prime}e$,
\begin{eqnarray}
&&d\sigma_{RR}\sim\bar{v}|M_{RR}|^2\nonumber\\
&&\hspace{30pt}\sim\frac{2}{c^2}[(1-v)^2+c^4)]\\
&&d\sigma_{LR}\sim\bar{v}|M_{LR}|^2\nonumber\\
&&\hspace{30pt}\sim\frac{1}{c^2}[2v^2s^4]\\
&&d\sigma_{+R}\sim\bar{v}|M_{+R}|^2\nonumber\\
&&\hspace{30pt}\sim 2v^2s^2
\end{eqnarray}

Fig.\ref{evcs} illustrated explicit shape of the $v$ and $\theta$
dependence of the cross sections. From the figure, we can easily
see that, in both the $R\to R$ and $L\to R$ processes, large angle
scattering cross sections dominate over the little angle ones.
While, in the $+\to R$ process, the cross sections depend on the
scattering angle weakly. As we will indicate in the following, in
pion VCS, the same properties of the polarized cross sections
persist.

\begin{figure}
\begin{center}
\begin{minipage}[b]{1.0\textwidth}
\scalebox{0.42}[0.65]{\includegraphics*[-7pt,299pt][380pt,572pt]{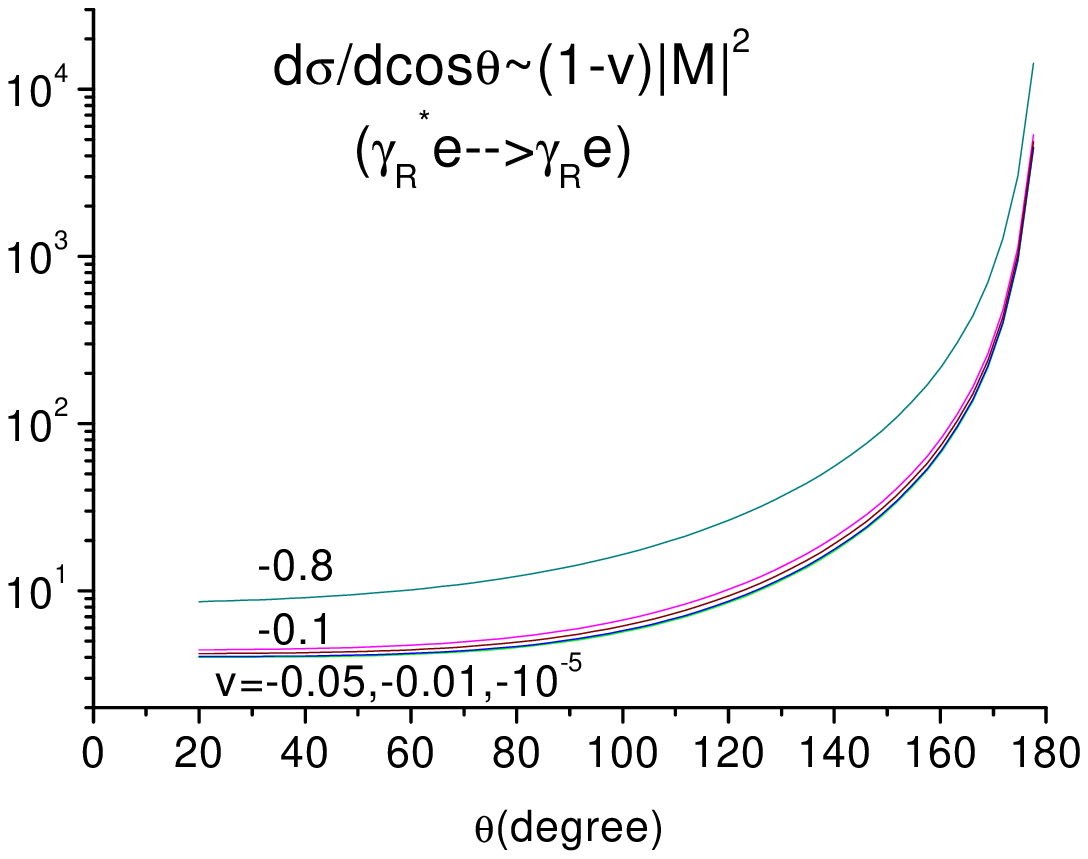}}
\scalebox{0.42}[0.65]{\includegraphics*[48pt,299pt][380pt,550pt]{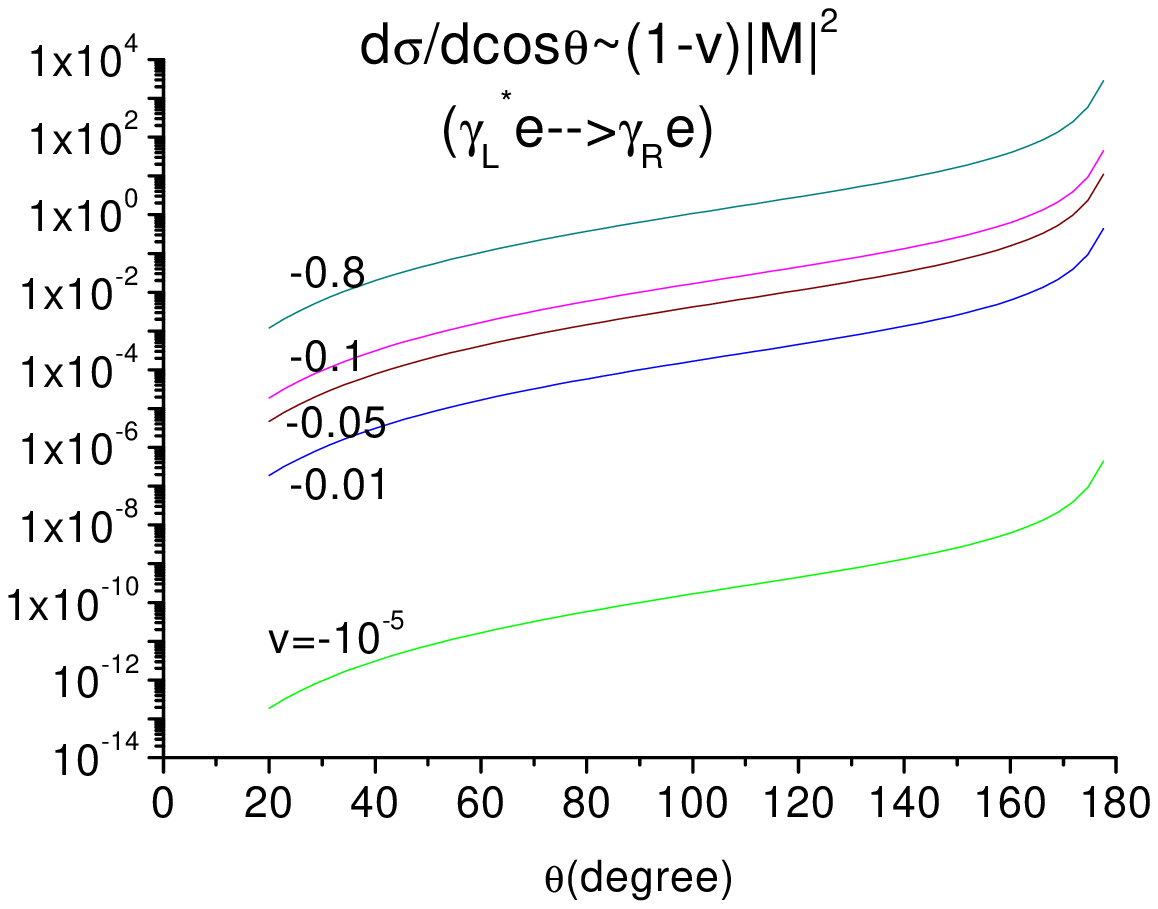}}
\scalebox{0.42}[0.65]{\includegraphics*[48pt,299pt][380pt,585pt]{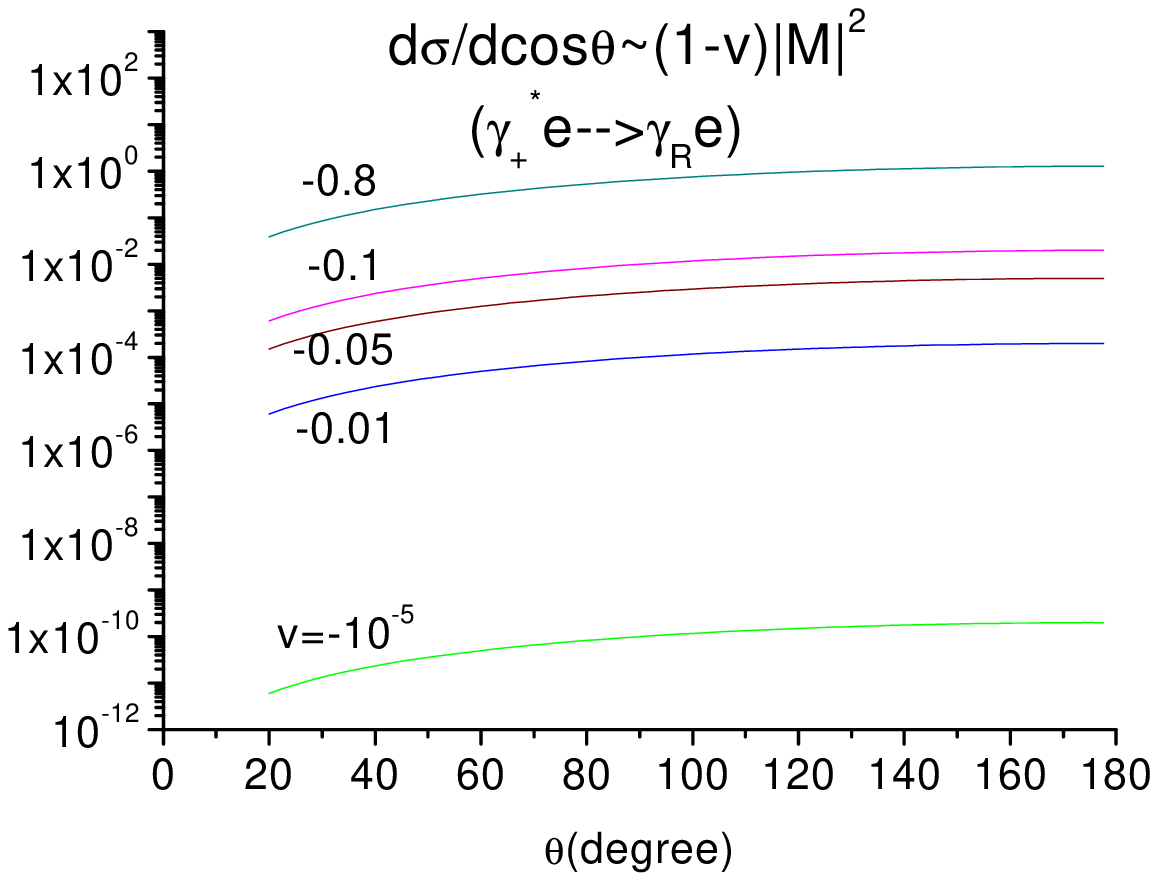}}
\end{minipage}
\hspace{1cm}
\begin{minipage}[c]{0.8\textwidth}
\caption{\small VCS on the unpolarized electron.}\label{evcs}
\end{minipage}
\end{center}
\end{figure}

\section{Numerical Integrations and Results}\label{numerical-res-sec}

We perform numerical integrations and give results for the cross
sections as well as corresponding phases for different polarized
processes in this section.

With eqs.(\ref{factor}), (\ref{da}), (\ref{sumdiagram}) and table
1-3, and using the following relation\cite{Dixon}
\begin{equation}
S^3\frac{d\sigma_{\epsilon\epsilon^\prime}(\theta)}{d\cos\theta}
=\frac{1}{2}\bar{v}S^4\frac{d\sigma_{\epsilon\epsilon^\prime}(\theta)}{dt}
=\bar{v} \frac{S^2}{32\pi}|M_{\epsilon\epsilon^\prime}|^2,
\end{equation}
and the principle integration formula,
\begin{eqnarray}
&&\int dxdy\frac{f(x,y)}{x-a(y)+i\epsilon}= \int
dxdyP\frac{f(x,y)}{(x-a)}-i\pi\int_0^1dxdyf(x,y)\delta(x-a)\;,
\nonumber\\
&&\int dxdyP\frac{f(x,y)}{x-a}=\int dxdyP\frac{f(x,y)-f(a,y)}{x-a}
+f(a,y)\log\frac{1-a}{a}\;,\label{principle}
\end{eqnarray}
we can get reliable numerical results for the cross sections and
corresponding phases for different polarization processes. One can
see that our treatment of the singular point appearing in the
numerical integration is the subtraction method. As was indicated
by \cite{Dixon,KN} in the similar calculation for proton Compton
scattering, different numerical treatments of the singular points
can give consistent results as long as it is applied
appropriately.

We get our numerical integrations by VEGAS program \cite{VEGAS}.
As did by \cite{Maina}, we take $\alpha_s=0.3$,
$\alpha_e=1/137.036$ in numerical computations. Noting the fact
that, $H$, thus $M$ varies with $1\over S$, we show the product
$S^3d\sigma_{\epsilon\epsilon^\prime}(\theta)\over d\cos\theta$
instead of $d\sigma_{\epsilon\epsilon^\prime(\theta)}\over
d\cos\theta$ in final results.

First, in fig.\ref{ten-or-five}, using distribution amplitude
$\phi_{p3}$, we compared the cross sections of the process
$\gamma^*_L\pi^\pm\to\gamma_R\pi^\pm$ in the following two cases:
(i). the unintegrated amplitude includes 10 independent diagrams;
(ii). the unintegrated amplitude only includes 5 independent
diagrams (the upper part of fig.\ref{Feynman-diagram}). Obviously,
the cross sections from the ten diagram contained amplitudes can
not be gained by simply multiplying a total factor on those from
the five diagram contained ones.

Second, in fig.\ref{pr}, we reconstructed the results of
\cite{Maina} for the cross section
$S^3\frac{d\sigma_{+R}}{d\cos\theta}$ of pion VCS using
distribution amplitudes $\phi_{p3}$. In the right part of this
figure, we studied the scattering angle dependence of the cross
section when $v\to0$, by letting
$v=-0.8,-0.1,-0.05,-10^{-2},-10^{-5},-10^{-8}$ in stead of
$v=-1.0,-0.75,-0.50,-0.25$, as was done in \cite{Maina}.

In fig.\ref{v-dep}, we illustrated the $v$ and $\theta$ dependence
of the cross section and corresponding phases for different
polarized processes, with distribution amplitude $\phi_{as}$. For
the $+\to R$ process, the cross section decreases with $v\to0$,
and it equals to zero when $v=0$. The scattering angle dependence
of it is very similar to that of the electron VCS. As in the
electron VCS, both for $L\to R$ and for $R\to R$ processes, large
angle scattering cross sections dominate. The difference is, in
the little scattering angle regions, the cross section decreases
as $v\to 0$, while in the large angle region, the trend reverses.
When $v\to 0$, the phase of $M_{LR}\rightarrow180^\circ$, and if
we redefine the domain of the phase angle, it can be set to 0.

In fig.\ref{mod-dep-rcmptn}, we compared distribution amplitudes
from different models and their effects on the physical
observables. We must admit that, relative to those given by
$\phi_{as}$ and $\phi_{cz}$, cross sections given by the
end-point-region suppressed distribution amplitudes $\phi_{bhl}$
and $\phi_{hs}$ suffer some suppressions. We know that, if the
very end point region of the distribution amplitudes has very
important contributions to the cross sections of physical
processes, pQCD is non-applicable in calculating this problem.
Now, our results indicate that this is not the case, so our
calculation is self-consistence. Of course, to reduce the
differences from distribution amplitudes, careful treatments of
the end point region and consideration of the higher order
corrections are necessary in further study.

To see this more clearly, we computed another six polarized cross
sections for virtual Compton scattering process in
fig.\ref{mod-v-dep}. The upper part of fig.\ref{mod-v-dep}
corresponds to the $v=-0.05$ case while the down part corresponds
to the $v=-0.8$ case. As in fig.\ref{mod-dep-rcmptn}, we can see
that the end-point-region suppressed distribution amplitudes still
do not give much-suppressed cross sections at most of the
scattering angles.

\section{Conclusions}

We recalculated pion VCS in pQCD in this paper, RCS is treated as
a limit case in our framework. Comparisons with existing
literatures and with electron VCS are made in the text. Our study
of different distribution amplitudes and their effects on the
cross sections and corresponding phases of the polarized processes
indicates that the behavior of the distribution amplitudes at the
very end point region does not have very strong effects on the
physical predictions, but careful treatments of the end point
region of distribution amplitudes are necessary in the further
investigations of this problem.

\section*{Acknowledgements}
The first author thanks very much to Professor Hsiang-nan Li for
his suggestion of studying an exclusive process as the beginning
of pQCD learning. We are greatly indebted  to the anonymous
referee for the kind and valuable instructions and suggestions.
This work is partially supported by the National Natural Science
Foundation of China.

\newpage

\begin{figure}
\begin{center}
\begin{minipage}[c]{1.0\textwidth}
\scalebox{0.65}[0.65]{\includegraphics*[40pt,299pt][380pt,572pt]{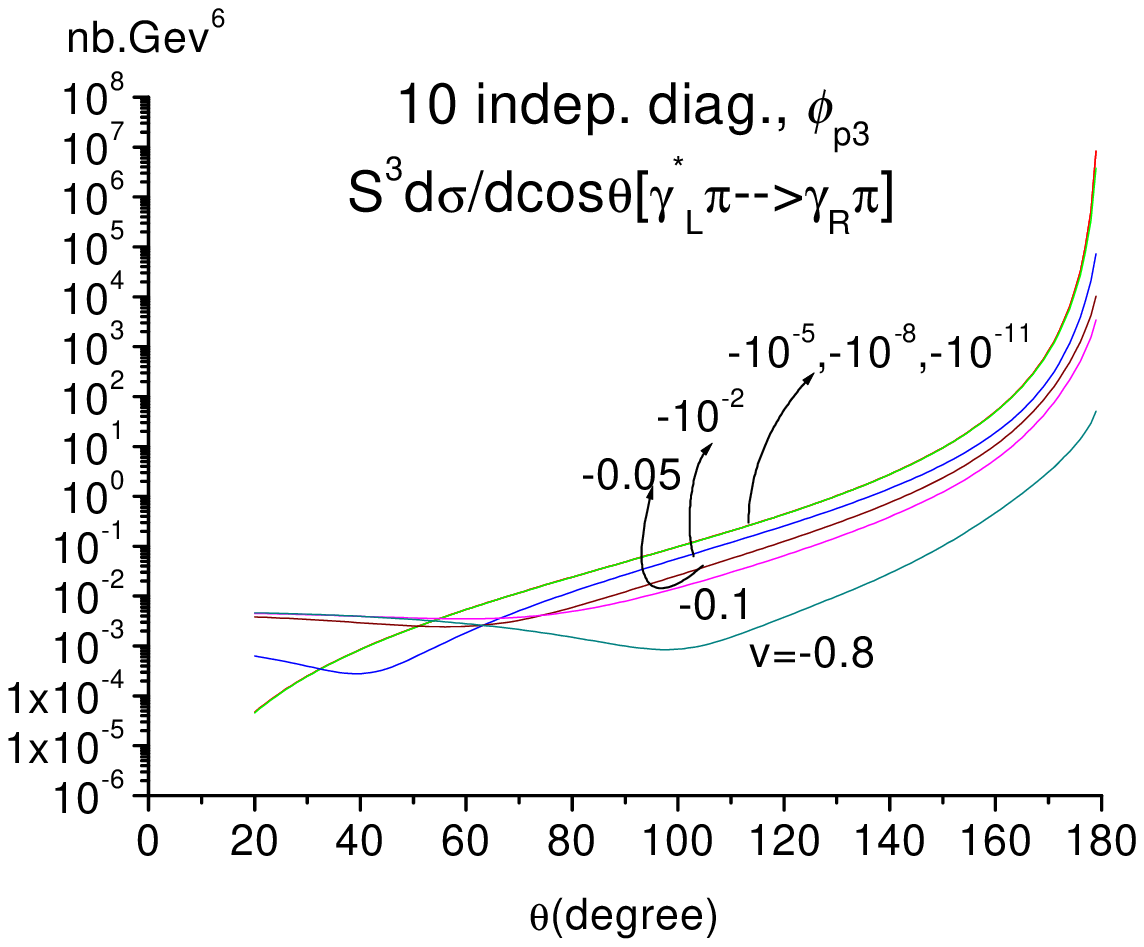}}
\scalebox{0.65}[0.65]{\includegraphics*[45pt,299pt][380pt,572pt]{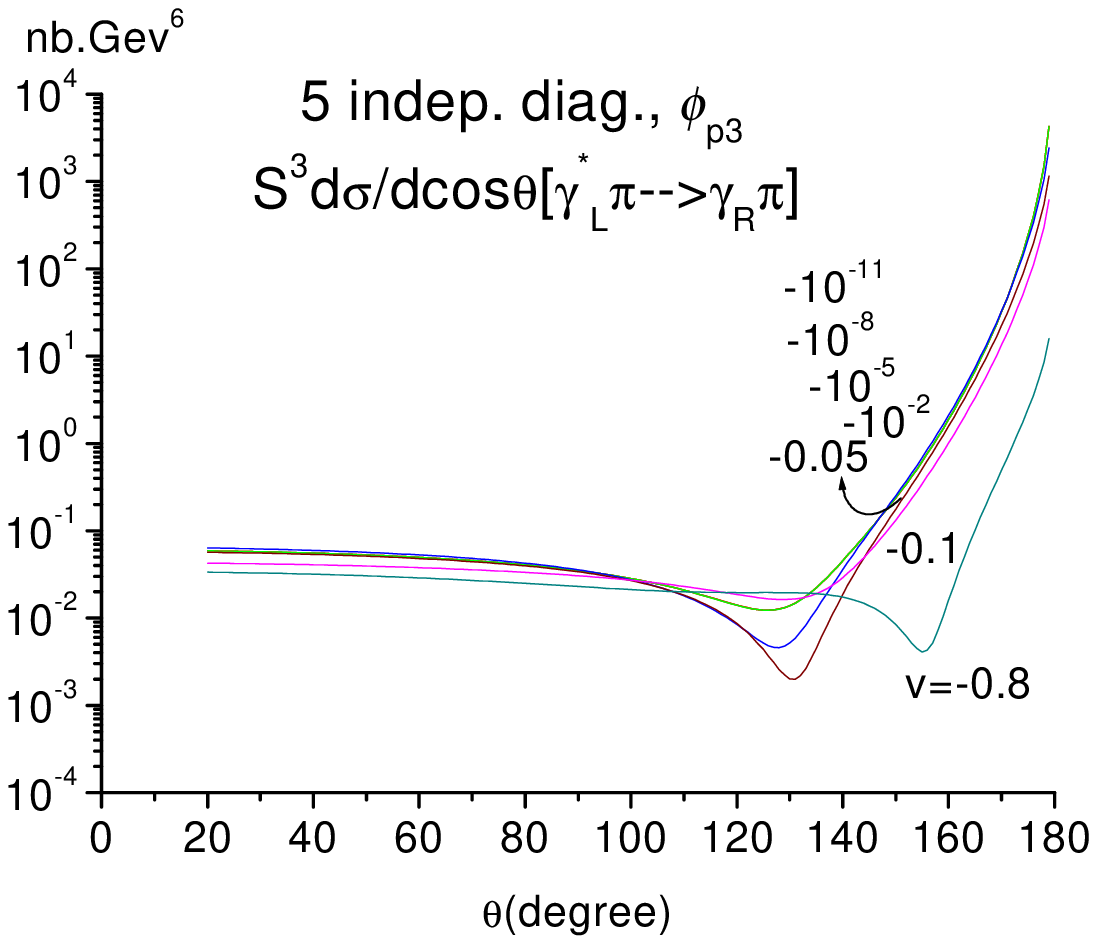}}
\end{minipage}
\hspace{1cm}
\begin{minipage}[c]{0.8\textwidth} \caption{\small
Left part, the cross section $S^3\frac{d\sigma_{LR}}{dcos\theta}$
when the hard amplitudes include ten independent diagrams; Right
part, the same quantity when the hard amplitudes only include five
independent diagrams.}\label{ten-or-five}
\end{minipage}
\end{center}
\end{figure}

\begin{figure}
\begin{center}
\begin{minipage}[c]{1.0\textwidth}
\scalebox{0.65}[0.65]{\includegraphics*[40pt,299pt][380pt,572pt]{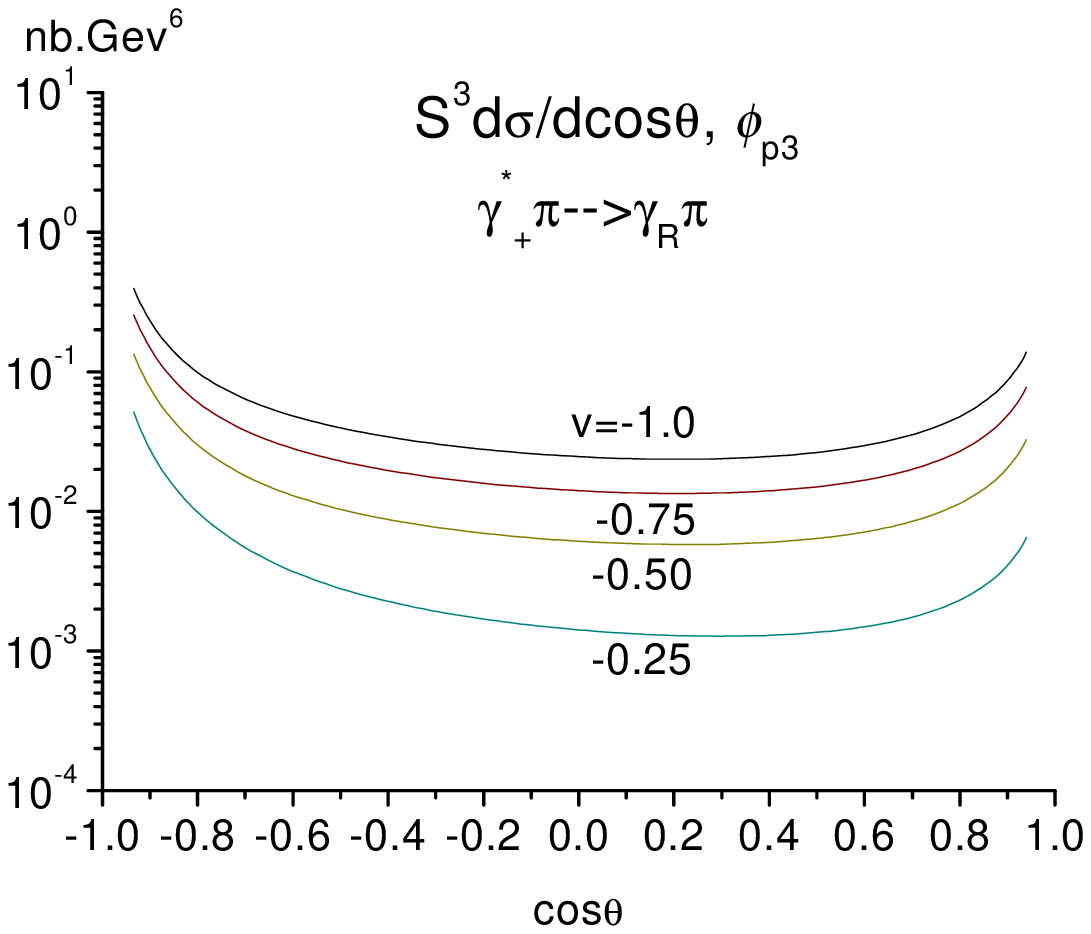}}
\scalebox{0.65}[0.65]{\includegraphics*[45pt,299pt][380pt,572pt]{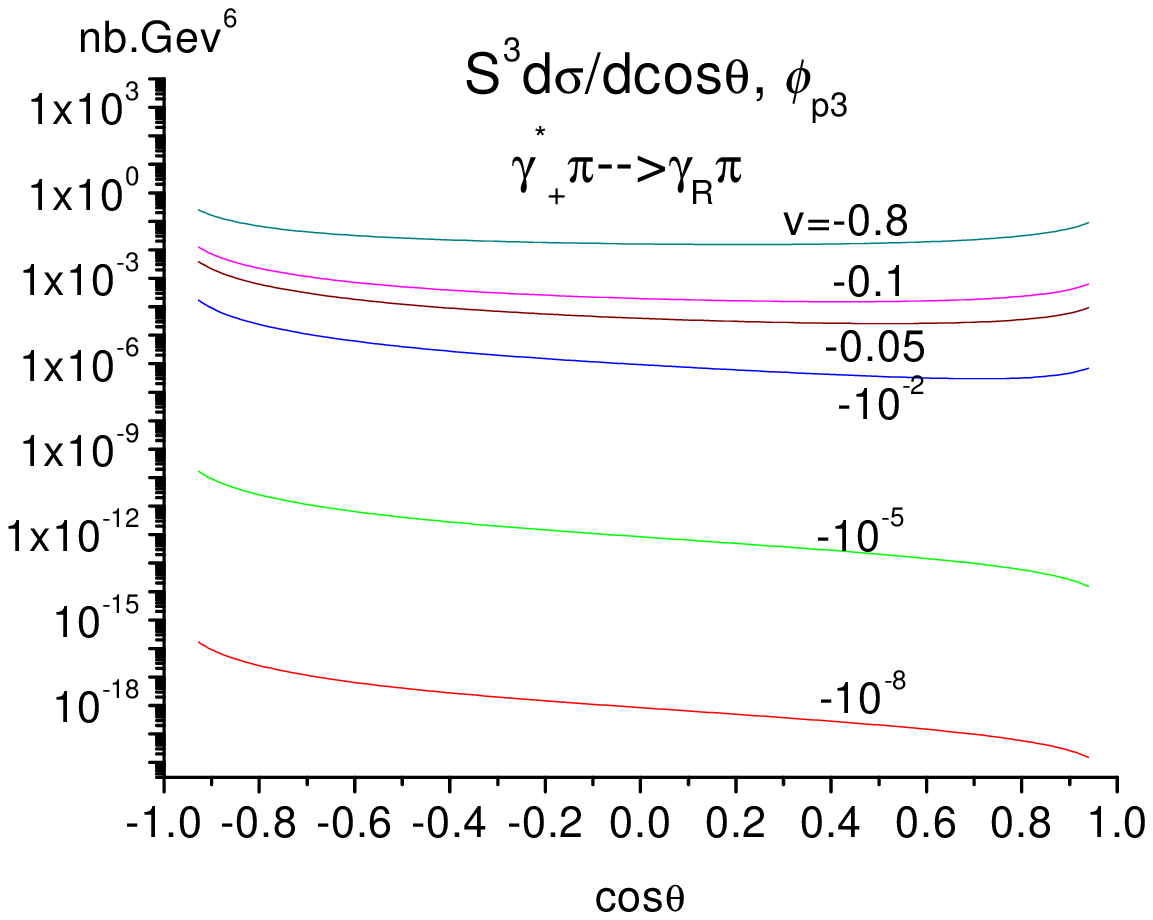}}
\end{minipage}
\hspace{1cm}
\begin{minipage}[c]{0.8\textwidth} \caption{\small
Left part, results reconstructed for
$S^3\frac{d\sigma_{+R}}{d\cos\theta}$ of \cite{Maina}. Input
photon virtuality approaches to zero in the way
$v=-1.0,-0.75,-0.50,-0.25$. Right part, our results for the
scattering angle dependence of the cross section when $v$ goes to
zero in the way
$v=-0.8,-0.1,-10^{-2},-10^{-5},-10^{-8}$.}\label{pr}
\end{minipage}
\end{center}
\end{figure}

\begin{figure}
\begin{center}
\begin{minipage}[c]{1.0\textwidth}
\scalebox{0.43}[0.65]{\includegraphics*[3pt,299pt][380pt,572pt]{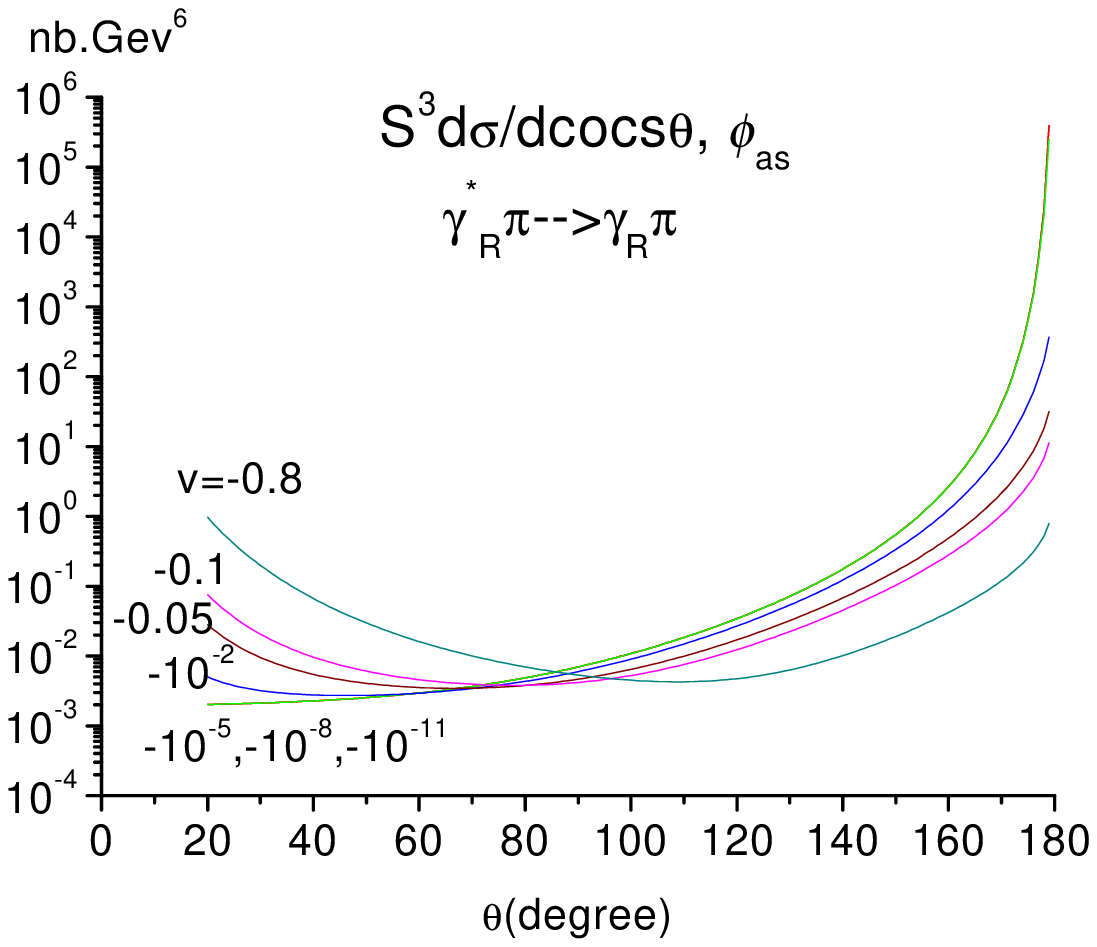}}
\scalebox{0.43}[0.65]{\includegraphics*[56pt,299pt][380pt,572pt]{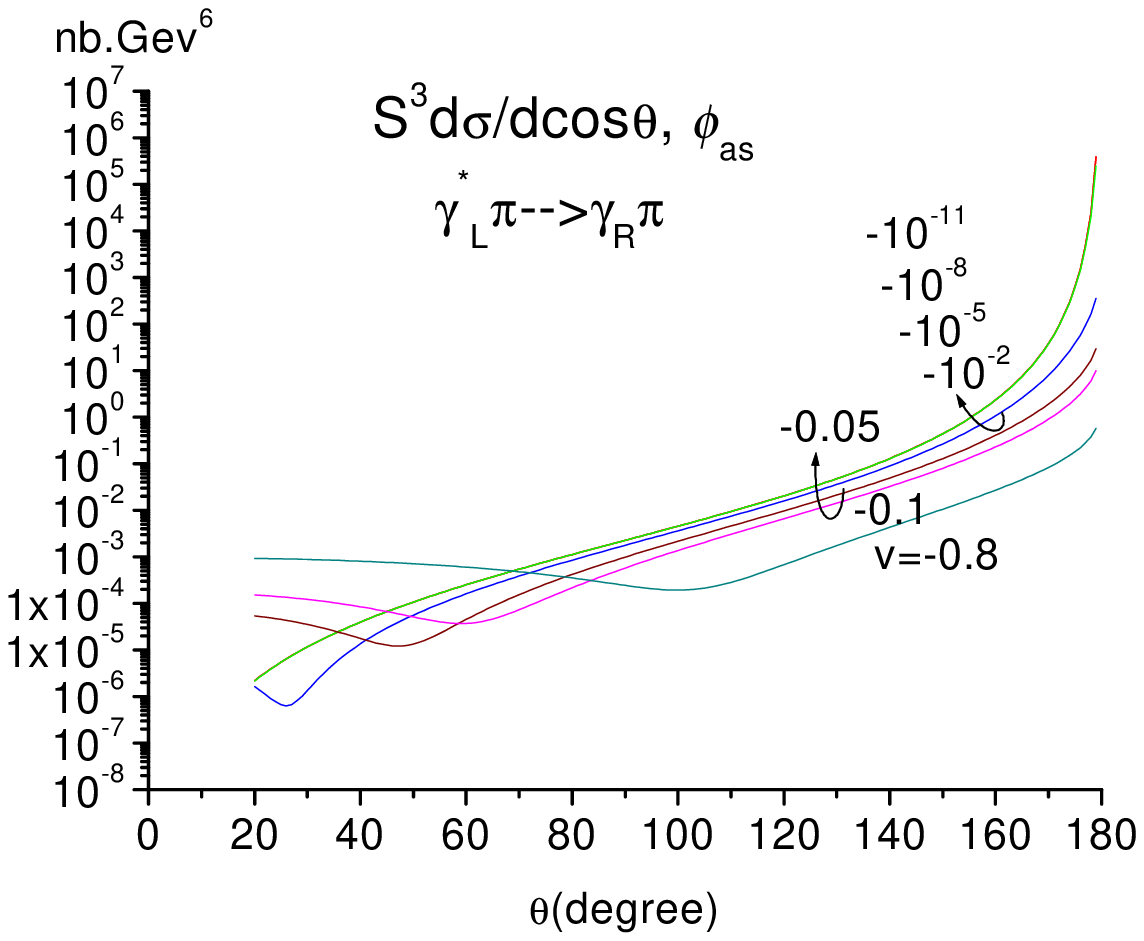}}
\scalebox{0.43}[0.65]{\includegraphics*[56pt,299pt][380pt,572pt]{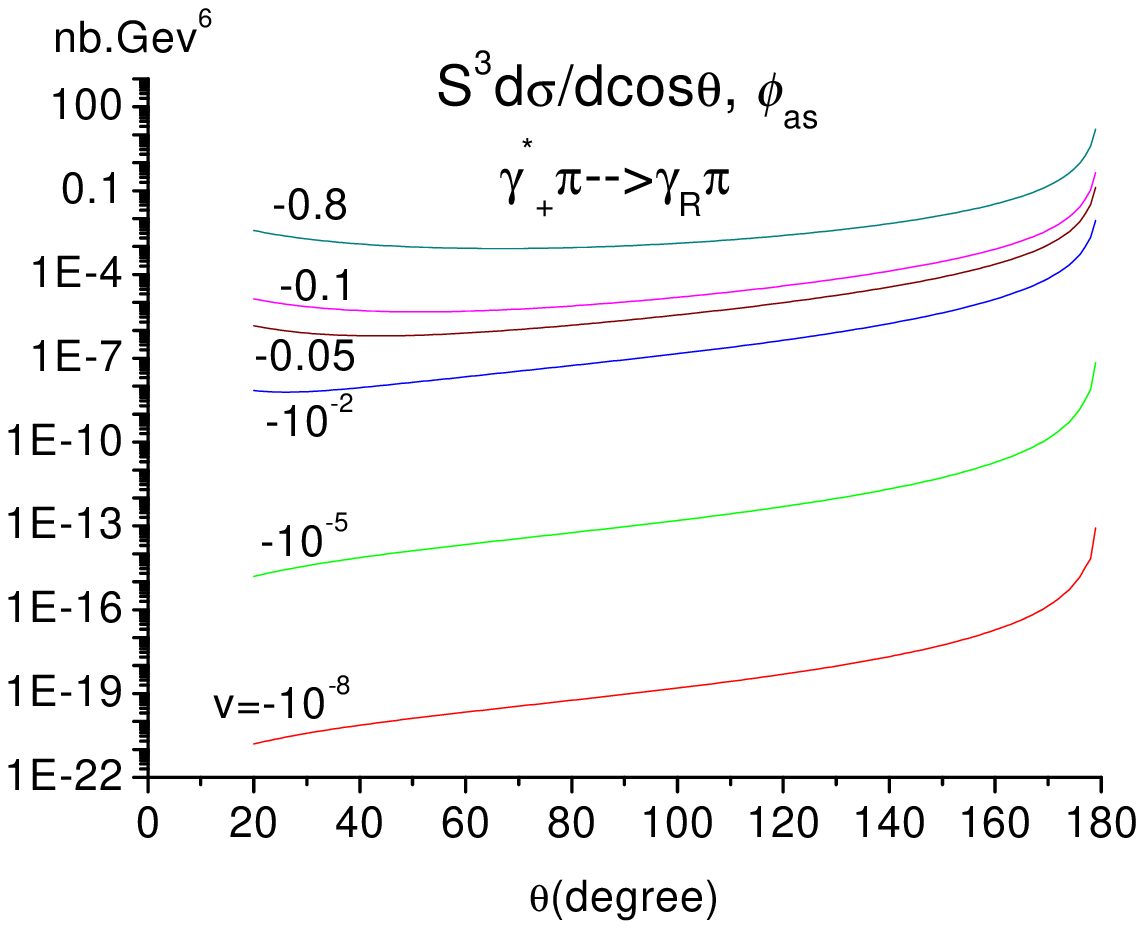}}\\
\scalebox{0.43}[0.65]{\includegraphics*[3pt,299pt][380pt,572pt]{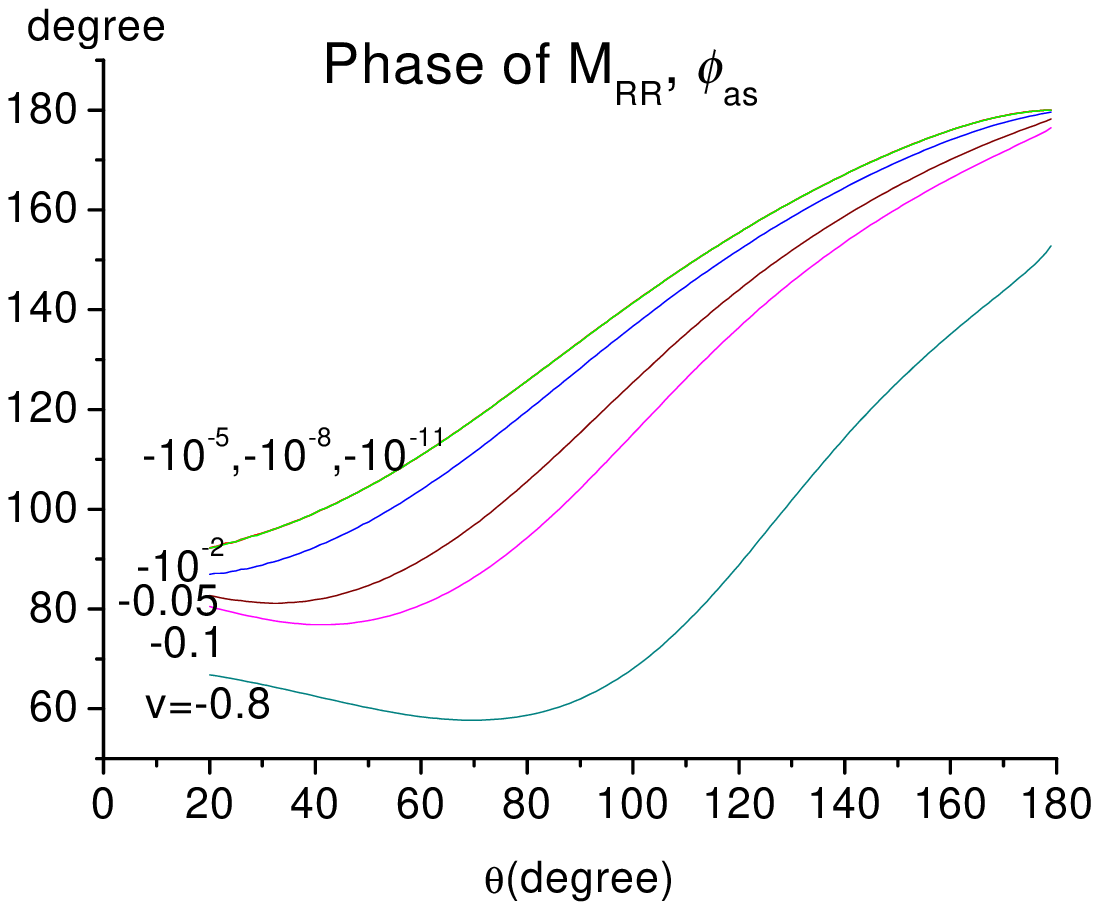}}
\scalebox{0.43}[0.65]{\includegraphics*[56pt,299pt][380pt,572pt]{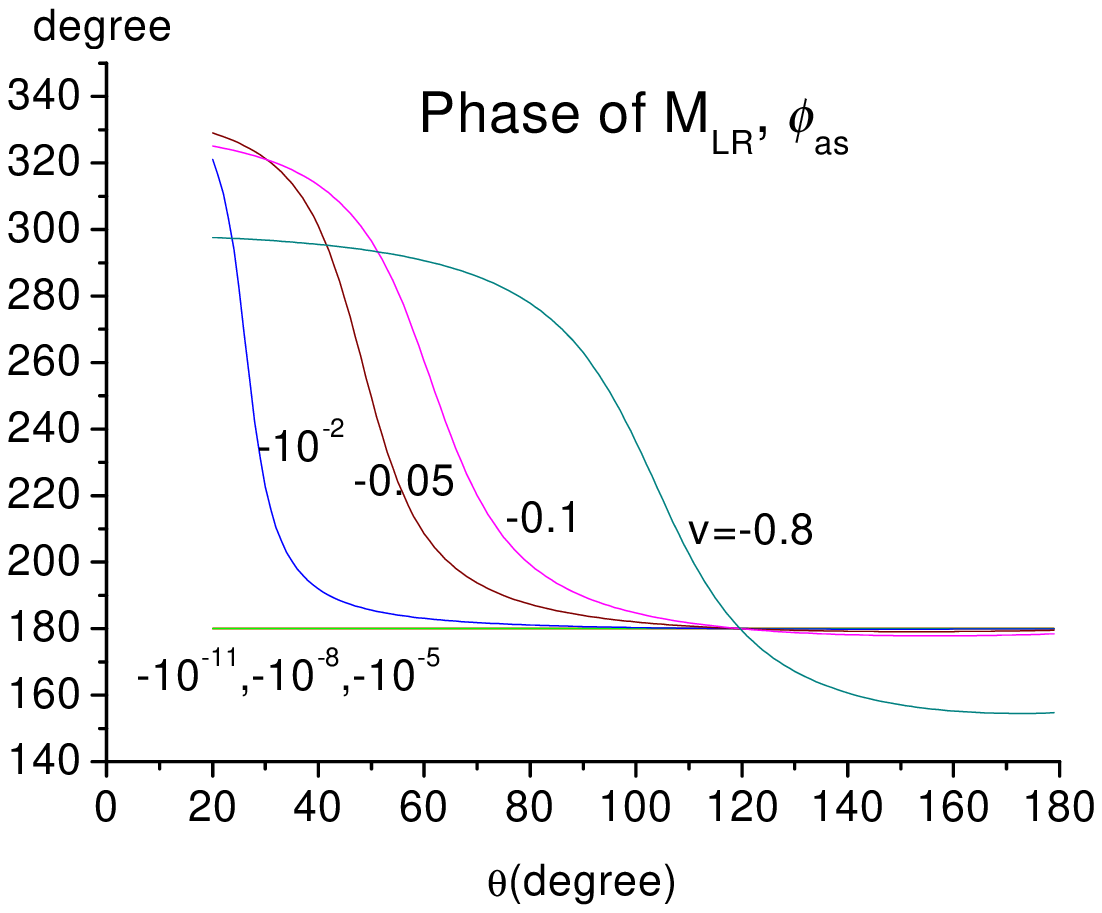}}
\scalebox{0.43}[0.65]{\includegraphics*[48pt,299pt][380pt,572pt]{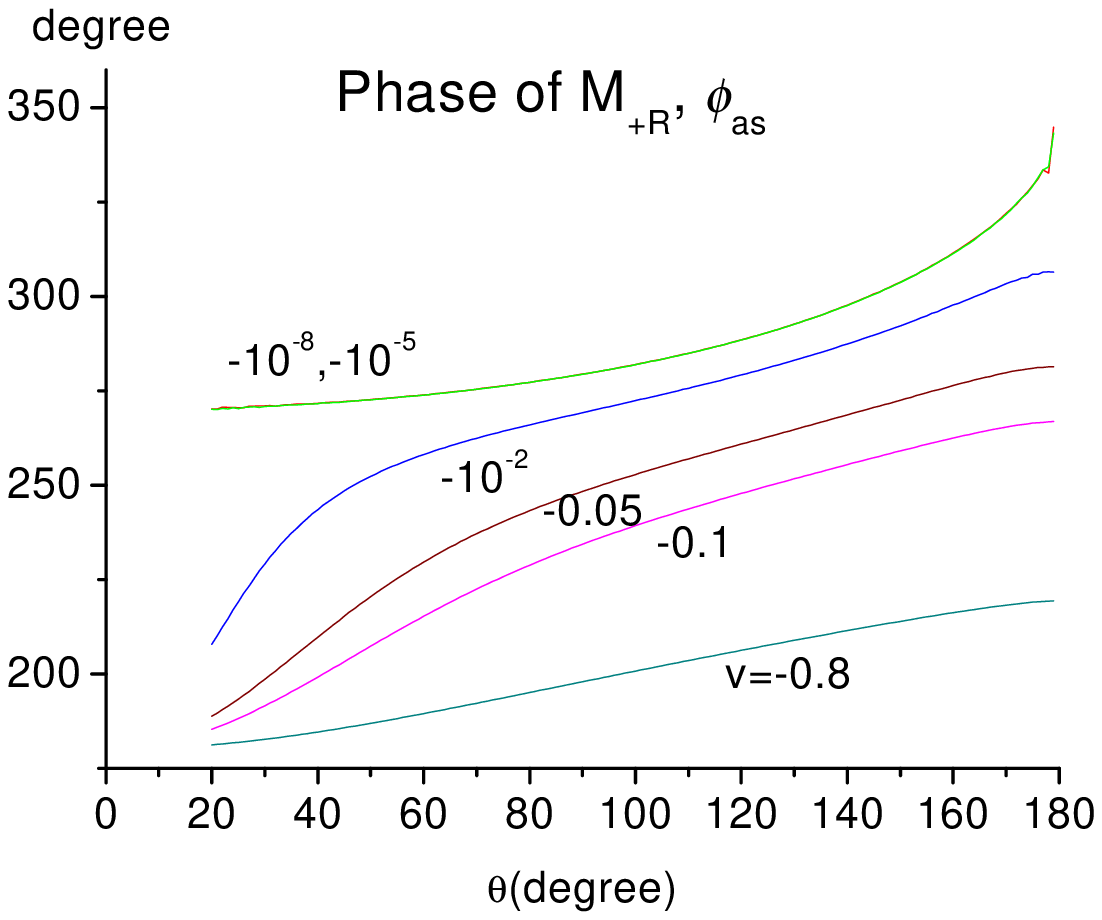}}
\end{minipage}
\hspace{1cm}
\begin{minipage}[c]{0.8\textwidth} \caption{\small
Effects of input photon virtuality on the polarized cross section
and corresponding phases. The distribution amplitude involved is
$\phi_{as}$. As $v\to 0$, the phase of $M_{LR} \to 180^\circ$
asymptotically.}\label{v-dep}
\end{minipage}
\end{center}
\end{figure}

\begin{figure}
\begin{center}
\begin{minipage}[b]{1.0\textwidth}
\scalebox{0.43}[0.65]{\includegraphics*[40pt,289pt][380pt,572pt]{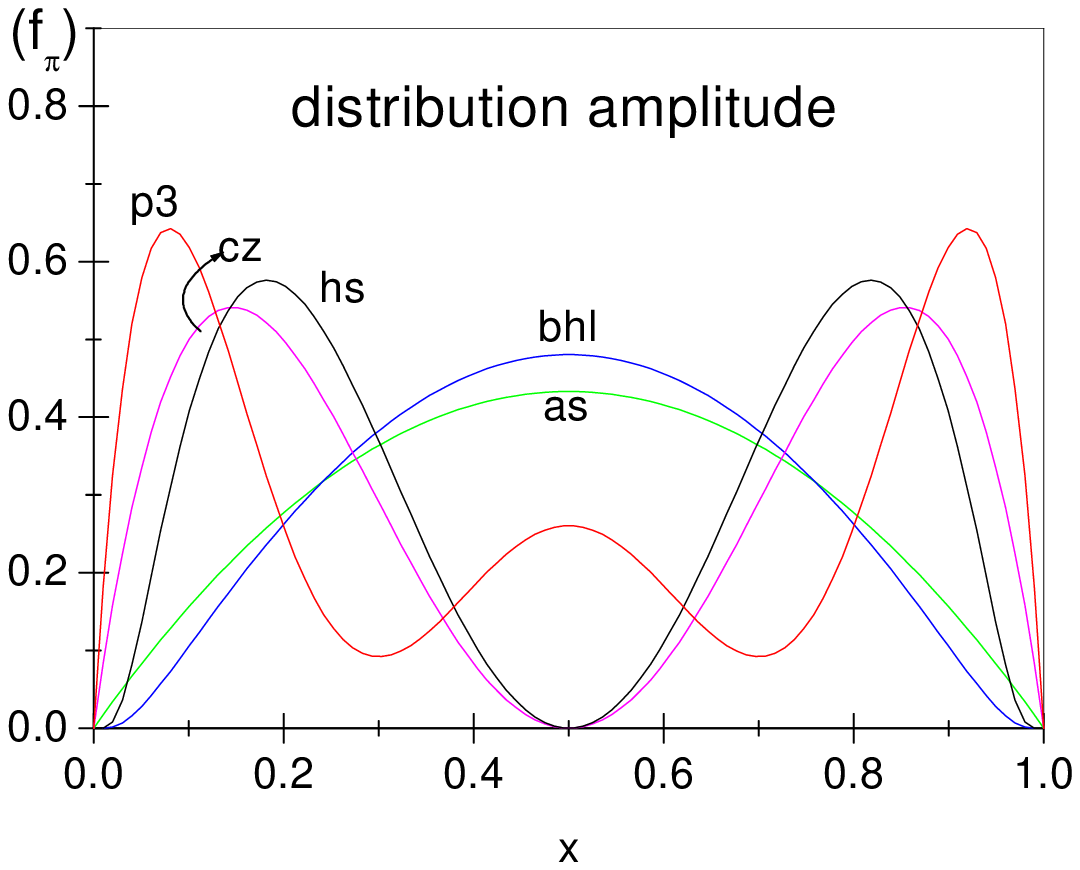}}
\scalebox{0.44}[0.65]{\includegraphics*[40pt,299pt][380pt,572pt]{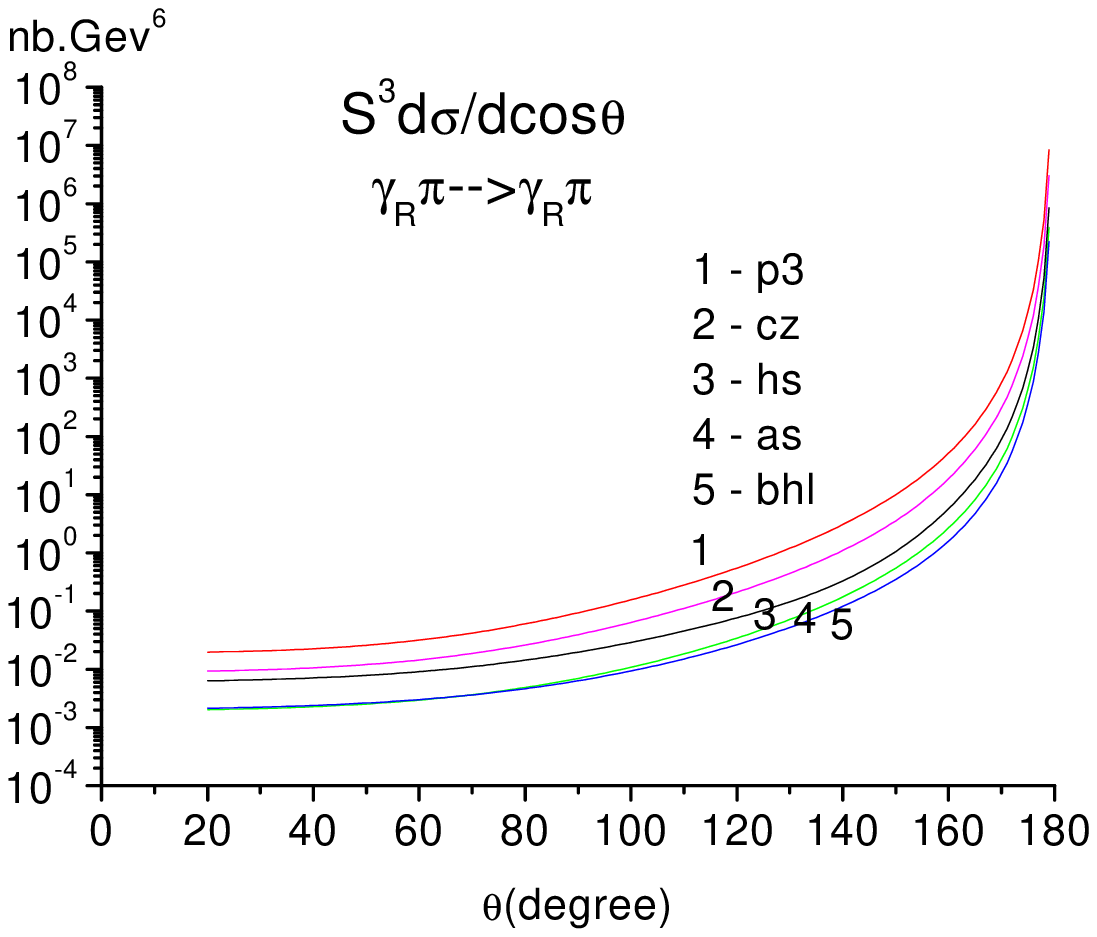}}
\scalebox{0.44}[0.65]{\includegraphics*[40pt,299pt][380pt,572pt]{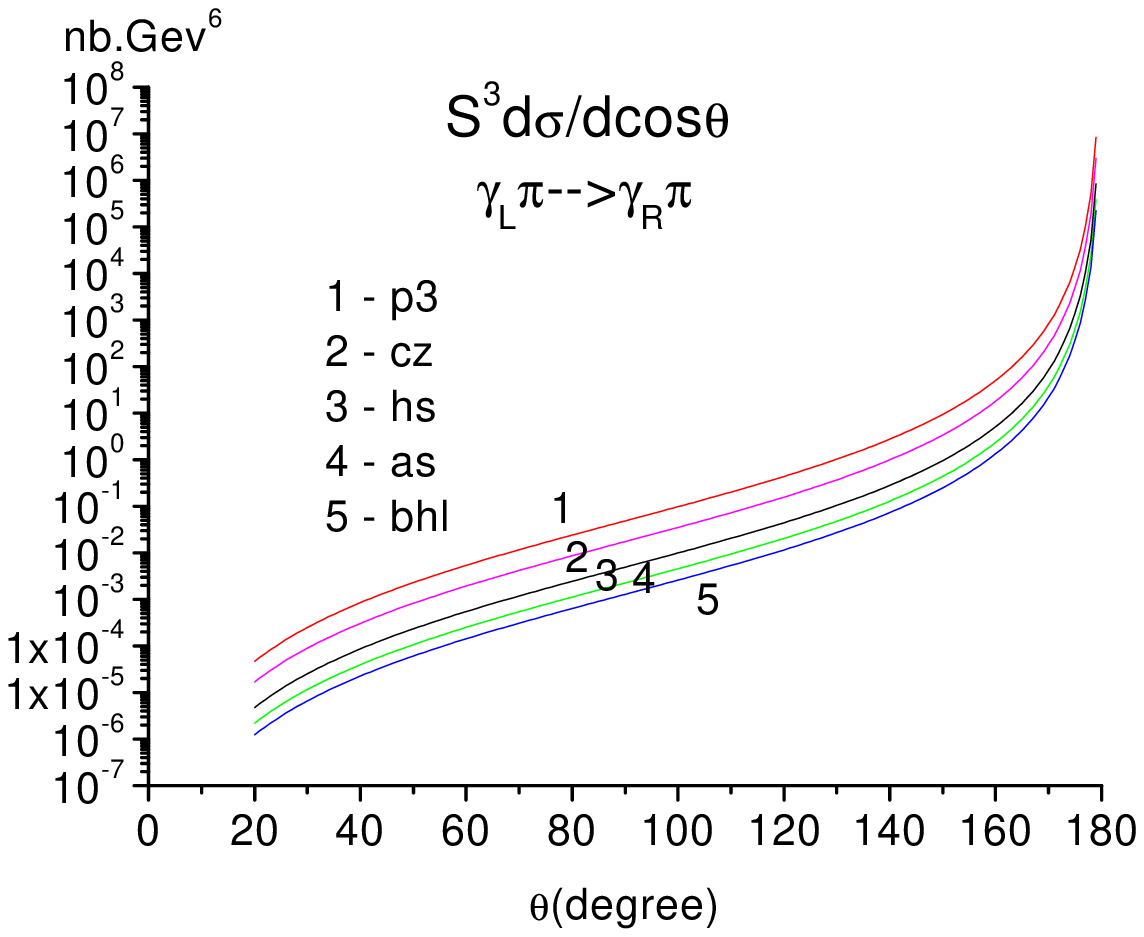}}\\
\scalebox{0.44}[0.65]{\includegraphics*[40pt,299pt][380pt,572pt]{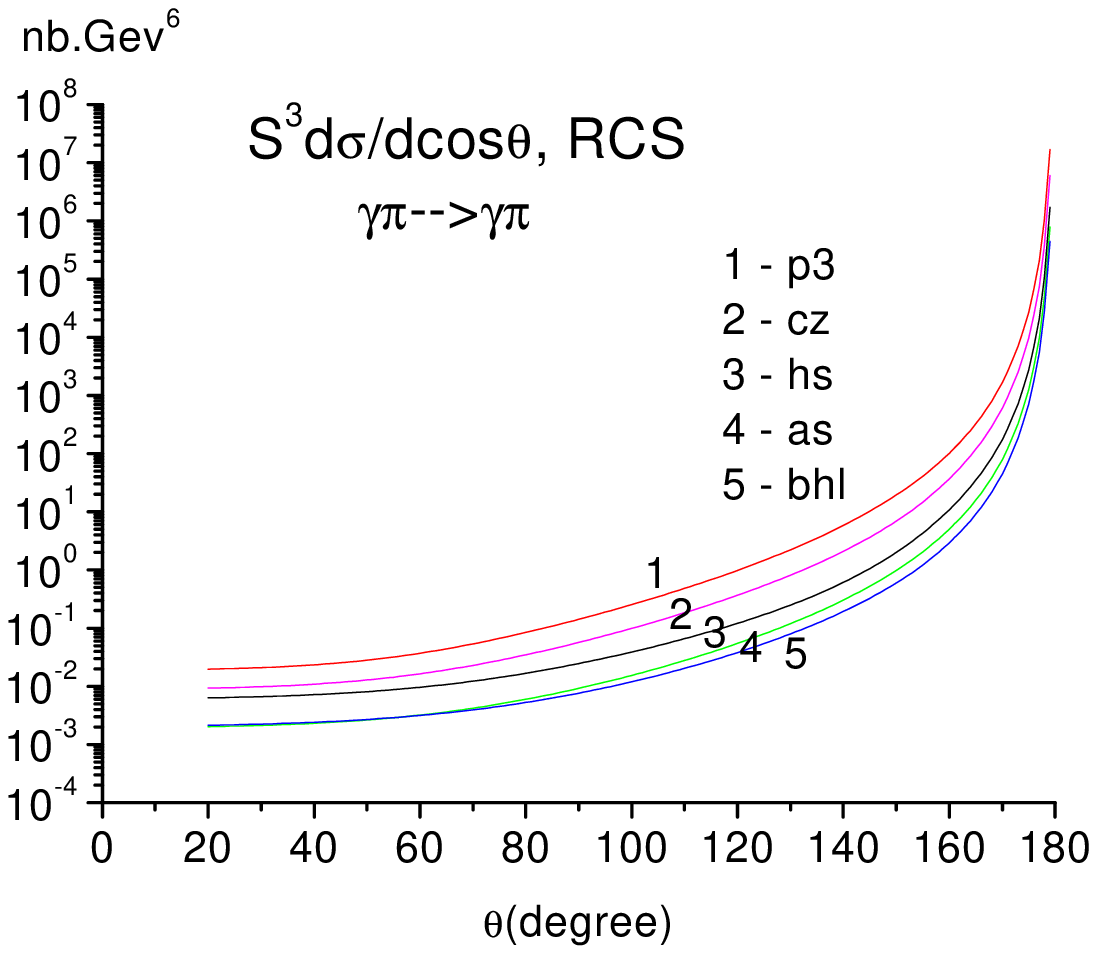}}
\scalebox{0.44}[0.65]{\includegraphics*[40pt,299pt][380pt,572pt]{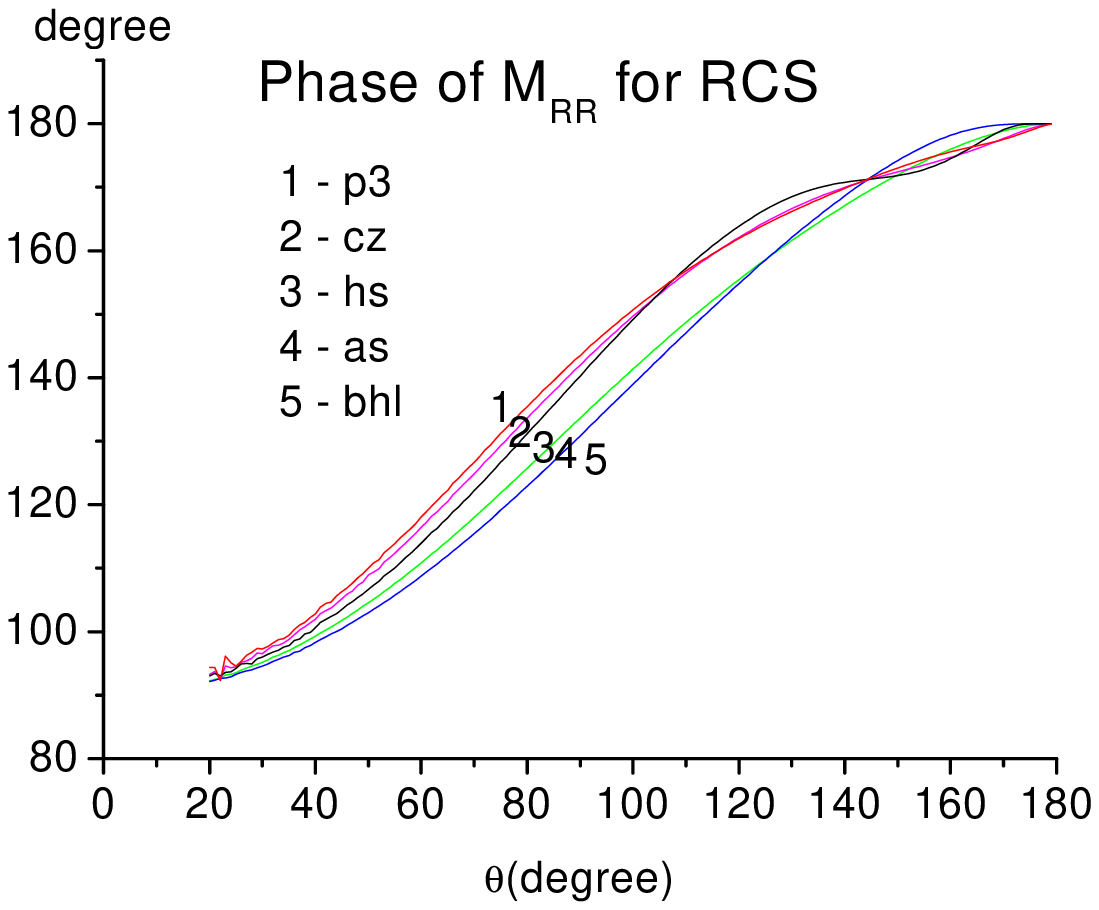}}
\scalebox{0.44}[0.65]{\includegraphics*[40pt,299pt][380pt,572pt]{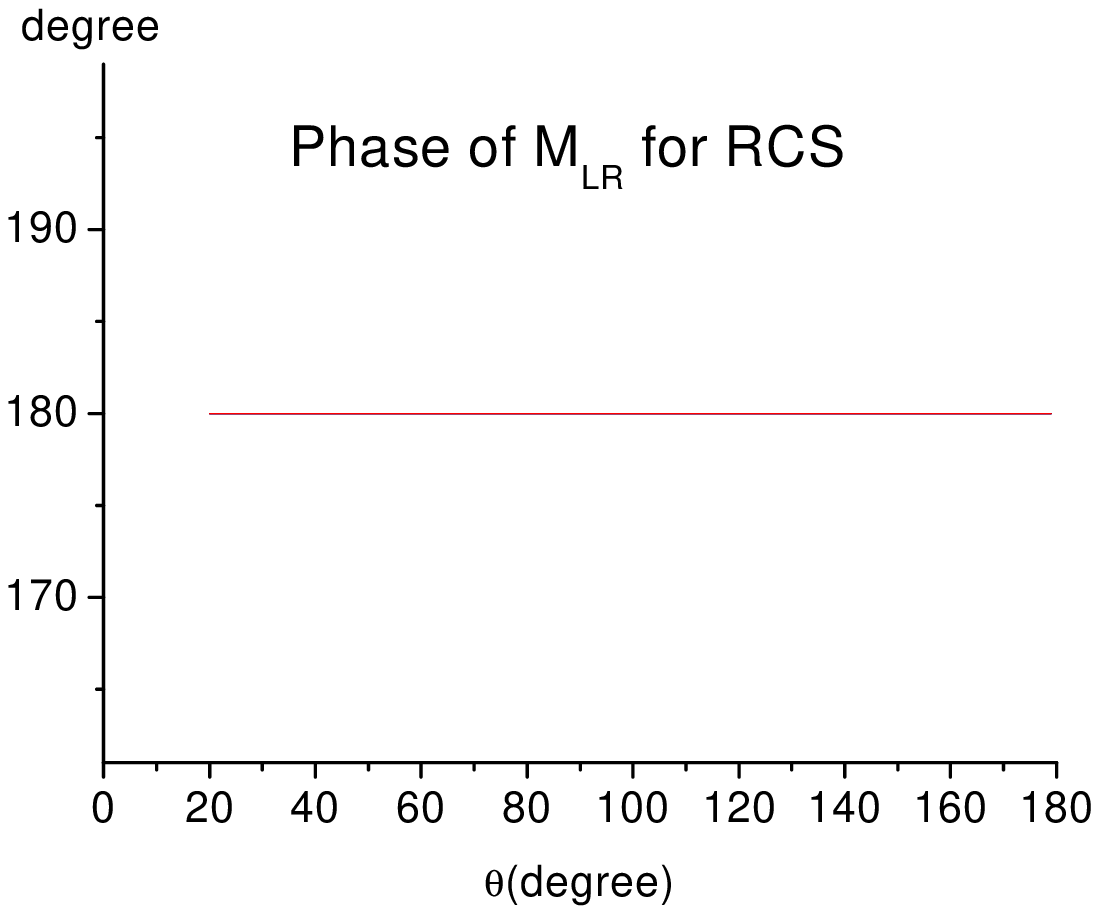}}
\end{minipage}
\begin{minipage}[c]{0.8\textwidth}
\caption{\small Distribution amplitudes and their effects on the
real Compton scattering process. Relative to $\phi_{as}$ and
$\phi_{cz}$, functions $\phi_{bhl}$ and $\phi_{hs}$ suppress the
end point region deeply, but the corresponding cross sections only
suffer little suppressions.}\label{mod-dep-rcmptn}
\end{minipage}
\end{center}
\end{figure}

\begin{figure}
\begin{center}
\begin{minipage}[b]{1.0\textwidth}
\scalebox{0.42}[0.65]{\includegraphics*[-7pt,299pt][380pt,572pt]{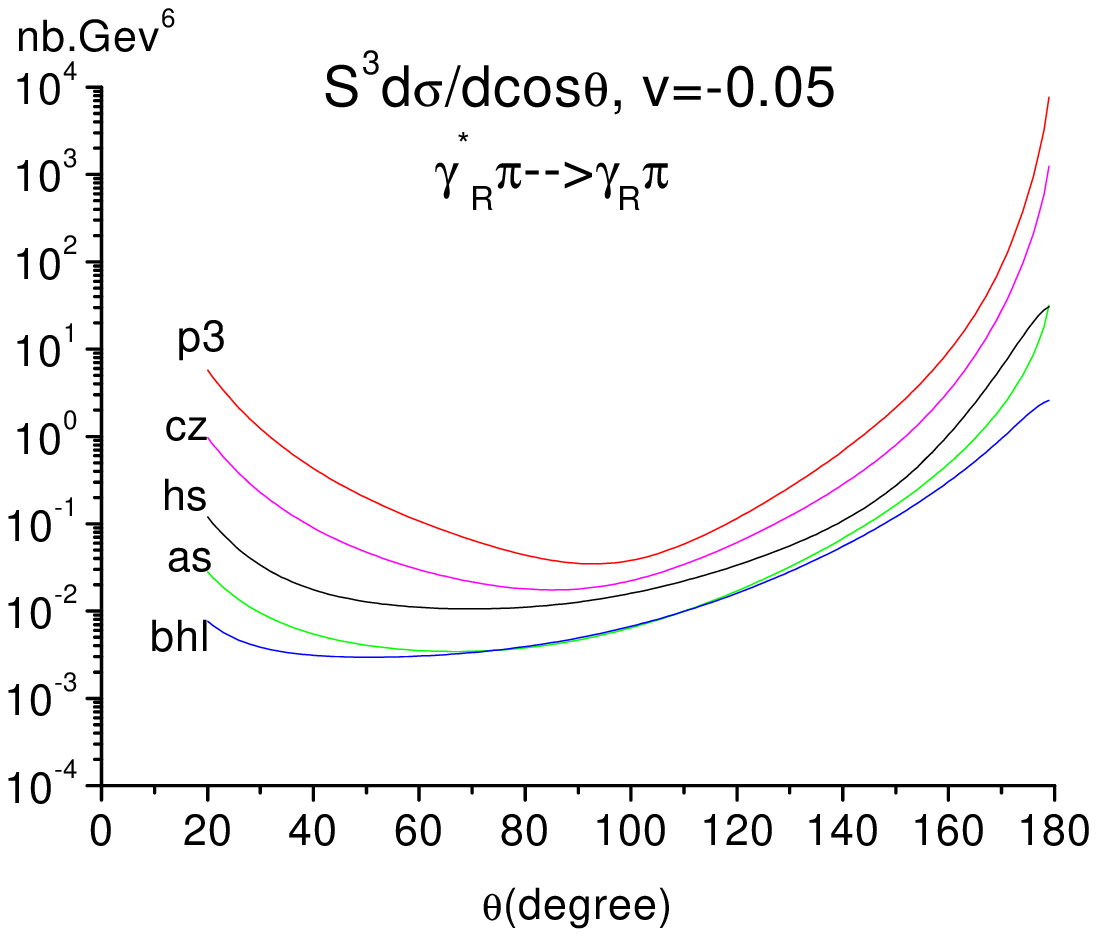}}
\scalebox{0.42}[0.65]{\includegraphics*[48pt,299pt][380pt,572pt]{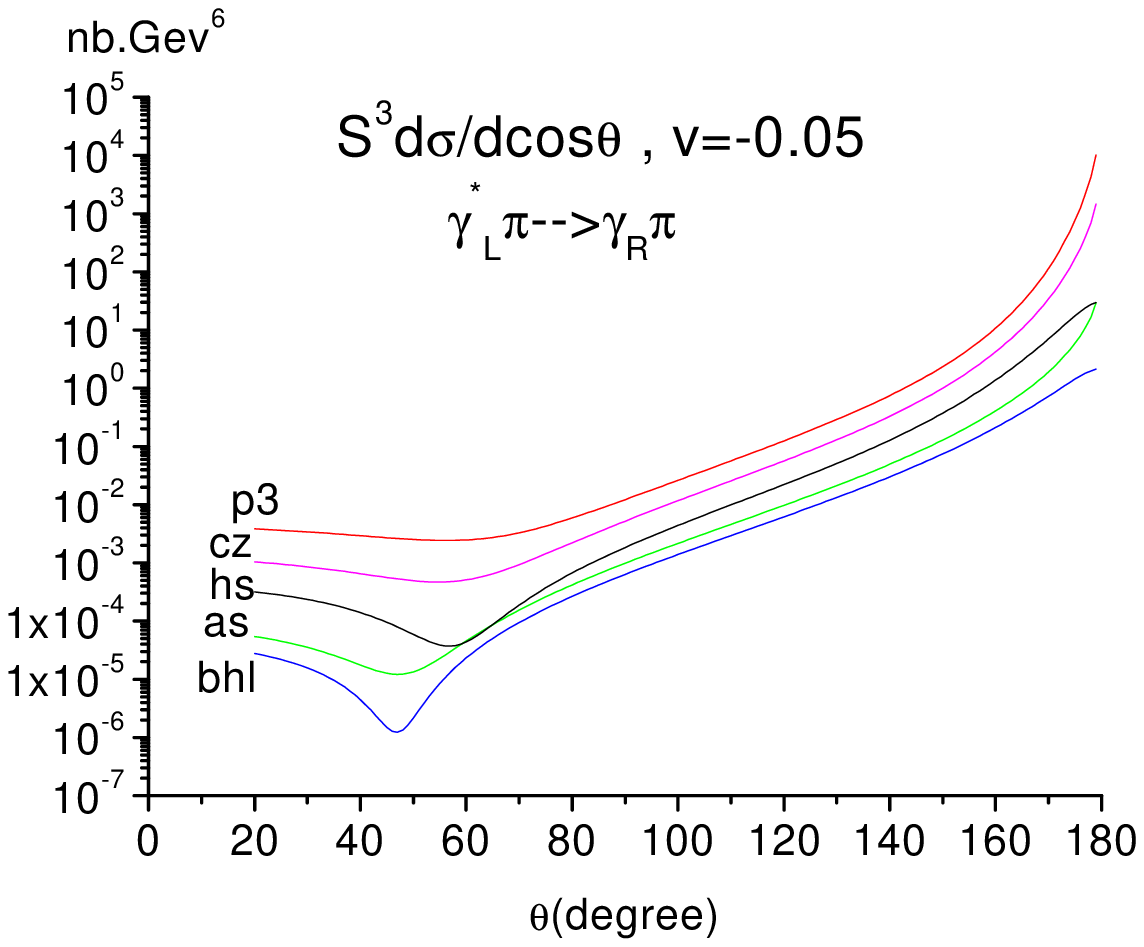}}
\scalebox{0.42}[0.65]{\includegraphics*[48pt,299pt][380pt,572pt]{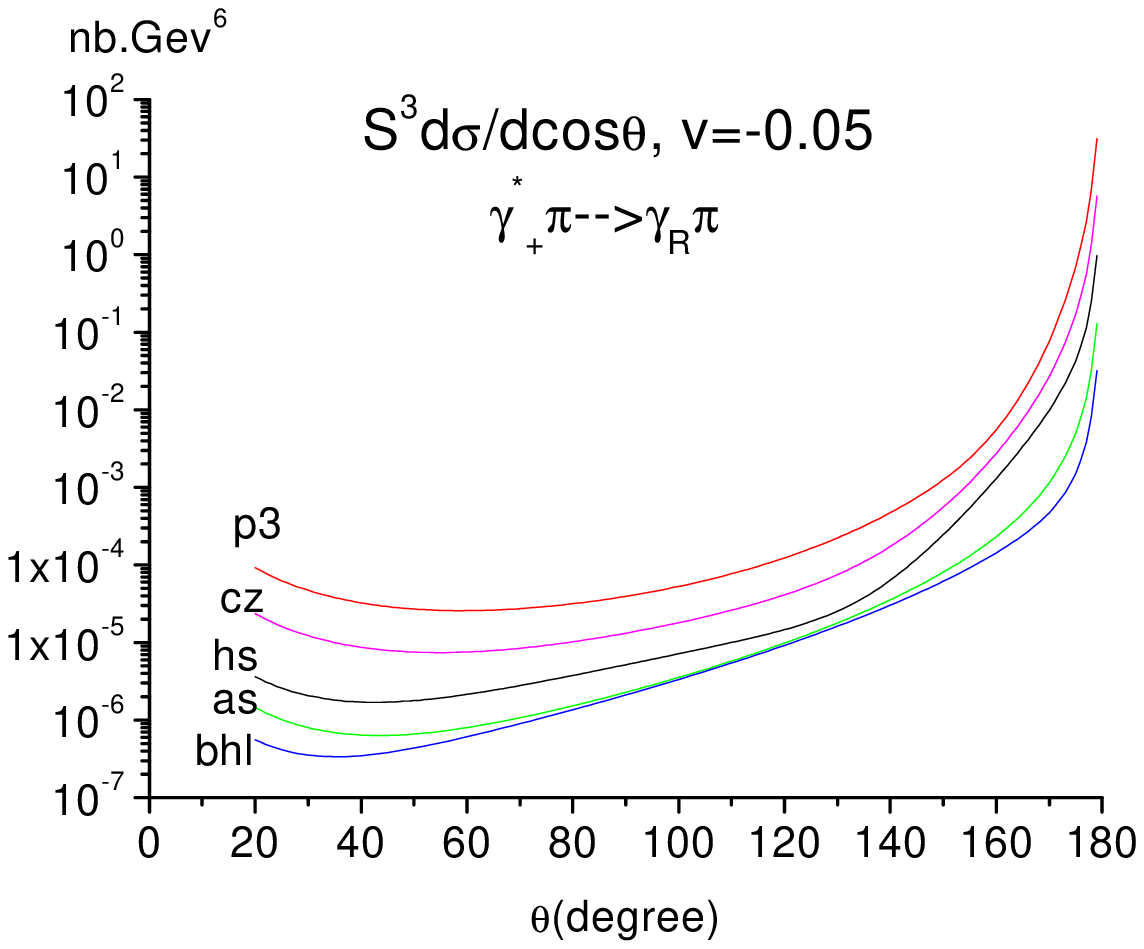}}\\
\scalebox{0.42}[0.65]{\includegraphics*[-7pt,299pt][380pt,572pt]{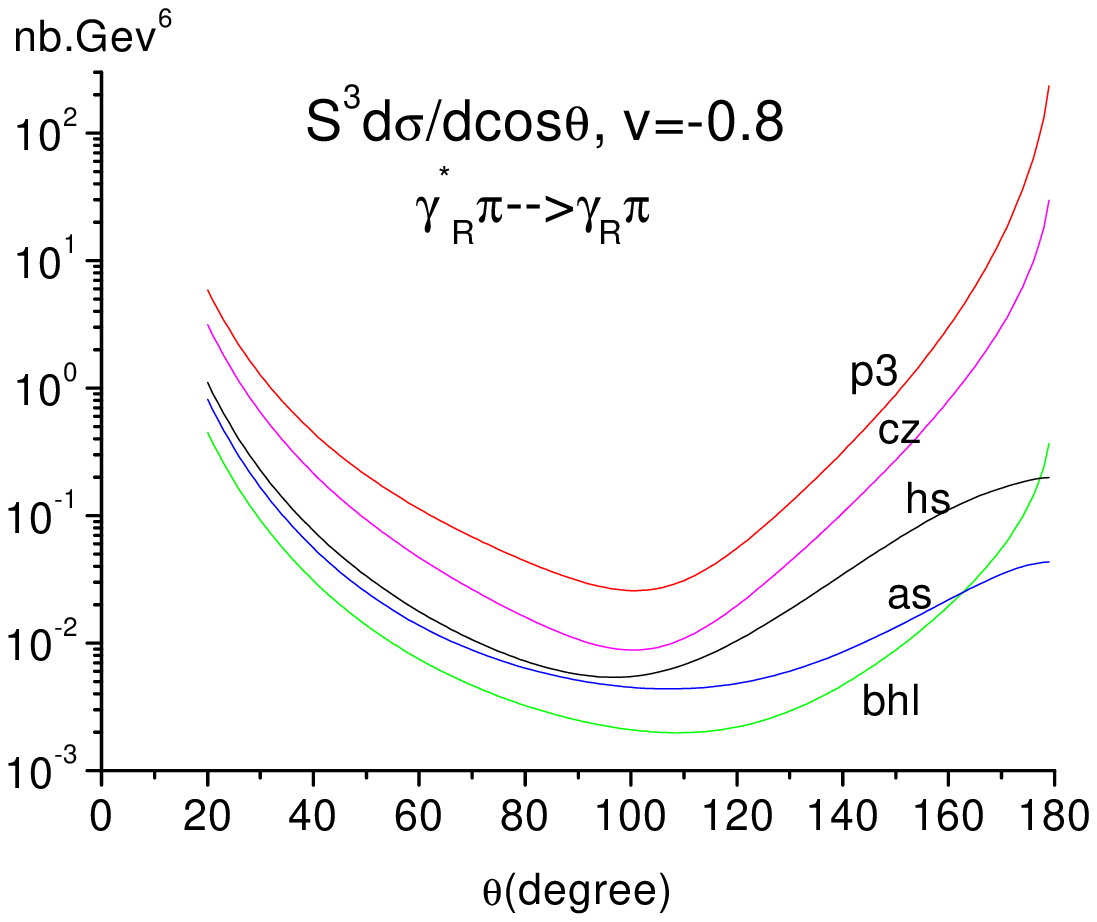}}
\scalebox{0.42}[0.65]{\includegraphics*[48pt,299pt][380pt,572pt]{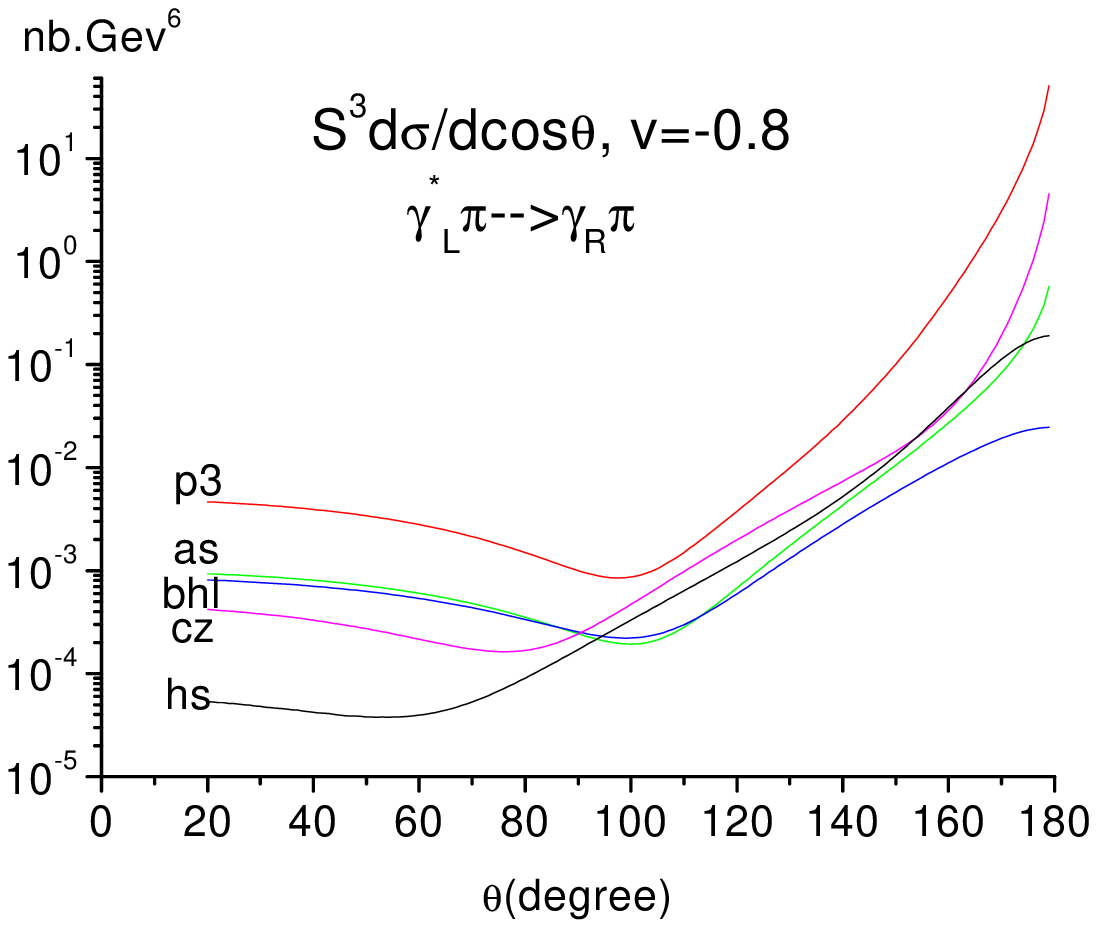}}
\scalebox{0.42}[0.65]{\includegraphics*[48pt,299pt][380pt,572pt]{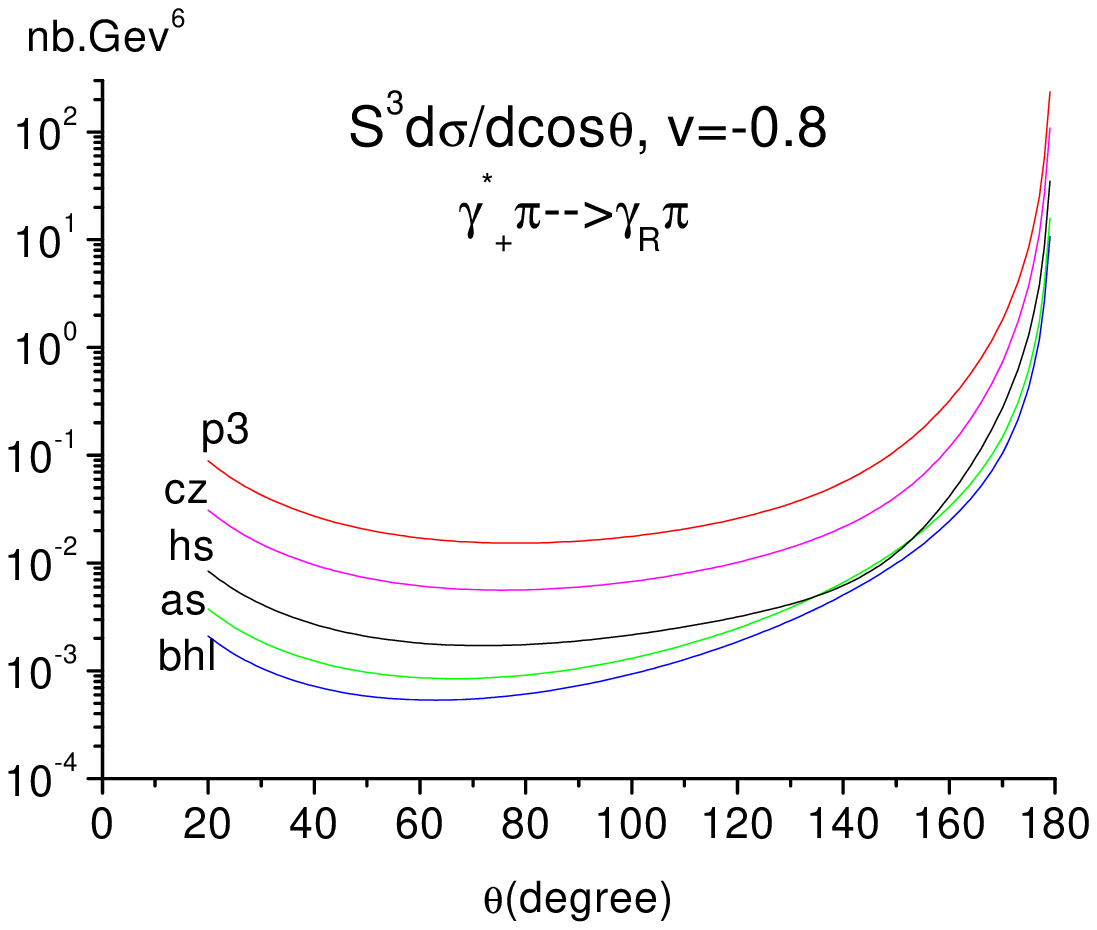}}
\end{minipage}
\hspace{1cm}
\begin{minipage}[c]{0.8\textwidth}
\caption{\small Further study of the effects of distribution
amplitudes on the virtual Compton scattering.}\label{mod-v-dep}
\end{minipage}
\end{center}
\end{figure}

\end{document}